%% file: main.tex
\documentclass[lettersize,journal]{IEEEtran}
\usepackage{amsmath,amsfonts}
\usepackage{algorithmic}
\usepackage{array}
\usepackage[caption=false,font=normalsize,labelfont=sf,textfont=sf]{subfig}
\usepackage{textcomp}
\usepackage{stfloats}
\usepackage{url}
\usepackage{verbatim}
\usepackage{graphicx}
\usepackage{bm}
\usepackage{amssymb}
\usepackage{algorithmic}
\usepackage{algorithm}
\usepackage{cite}
\usepackage{orcidlink}

\def\BibTeX{{\rm B\kern-.05em{\sc i\kern-.025em b}\kern-.08em
    T\kern-.1667em\lower.7ex\hbox{E}\kern-.125emX}}
\usepackage{balance}

\begin{document}
\bstctlcite{IEEEexample:BSTcontrol}

%%%%%%%%%%%%%%%%%%%%%%%%%%%%%%%%%%CopyrightNotice%%%%%%%%%%%%%%%%%%%%%%%%%%%%%%%%%
\onecolumn
\markboth{Journal of \LaTeX\ Class Files,~Vol.~18, No.~9, September~2020}%
{How to Use the IEEEtran \LaTeX \ Templates}

\vspace*{3cm}
{\Large IEEE Copyright Notice}
\vspace*{0.5cm}

Copyright 
$\copyright$ 2024 IEEE. Personal use of this material is permitted. Permission from IEEE must be obtained for all other uses, in any current or future media, including reprinting/republishing this material for advertising or promotional purposes, creating new collective works, for resale or redistribution to servers or lists, or reuse of any copyrighted component of this work in other works.

\vspace*{1cm}

\indent {\Large\textbf{Annealing-Assisted Column Generation \\ \indent for Inequality-Constrained Combinatorial Optimization Problems}}
\vspace*{0.5cm}

\indent Hiroshi Kanai \IEEEauthorrefmark{1} \\
\indent Masashi Yamashita\IEEEauthorrefmark{1} \\
\indent Kotaro Tanahashi\IEEEauthorrefmark{2} \\
\indent Shu Tanaka\IEEEauthorrefmark{1}\IEEEauthorrefmark{3}\IEEEauthorrefmark{4}\IEEEauthorrefmark{5}\IEEEauthorrefmark{6}

\vspace*{0.5cm}
\indent\IEEEauthorblockA{\IEEEauthorrefmark{1}\textit{\ The Graduate School of Science and Technology, Keio University}, Kanagawa 223-8522, Japan.} \\
\indent\IEEEauthorblockA{\IEEEauthorrefmark{2} \textit{Recruit Co., Ltd.}, Tokyo 100-6640, Japan.} \\
\indent\IEEEauthorblockA{\IEEEauthorrefmark{3} \textit{The Department of Applied Physics and Physico-Informatics, Keio University}, Kanagawa 223-8522, Japan.} \\
\indent\IEEEauthorblockA{\IEEEauthorrefmark{4} \textit{The Keio University Sustainable Quantum Artificial Intelligence Center (KSQAIC), Keio University}, \\ \indent \indent Tokyo 108-8345, Japan.} \\
\indent\IEEEauthorblockA{\IEEEauthorrefmark{5} \textit{The Human Biology-Microbiome-Quantum Research Center (WPI-Bio2Q), Keio University}, Tokyo 108-8345, Japan.} \\
\indent\IEEEauthorblockA{\IEEEauthorrefmark{6} \textit{The Green Computing Systems Research Organization (GCS), Waseda University}, Tokyo 162-0042, Japan.}

\vspace*{0.5cm}

\clearpage
%%%%%%%%%%%%%%%%%%%%%%%%%%%%%%%%%%CopyrightNotice%%%%%%%%%%%%%%%%%%%%%%%%%%%%%%%%%

\twocolumn
\title{Annealing-Assisted Column Generation \\ for Inequality-Constrained Combinatorial Optimization Problems}
\author{Hiroshi Kanai\,\orcidlink{0009-0003-6468-1036}, Masashi Yamashita\,\orcidlink{0009-0002-4654-7151}, Kotaro Tanahashi\,\orcidlink{0000-0001-8334-7285}, Shu Tanaka\,\orcidlink{0000-0002-0871-3836}, \textit{Member, IEEE}
\thanks{
This work was partially supported in part by JSPS KAKENHI (Grant Number JP23H05447), the Council for Science, Technology, and Innovation (CSTI) through the Cross-ministerial Strategic Innovation Promotion Program (SIP), ``Promoting the application of advanced quantum technology platforms to social issues'' (Funding agency: QST), JST (Grant Number JPMJPF2221), and JST CREST (Grant Number JPMJCR19K4).

Hiroshi Kanai and Masashi Yamashita are with the Graduate School of Science and Technology, Keio University, Kanagawa 223-8522, Japan (e-mail: kanai6939@keio.jp; mercy.under-mountain@keio.jp).

Kotaro Tanahashi is with Recruit Co., Ltd., Tokyo 100-6640, Japan (e-mail: tanahashi@r.recruit.co.jp).

Shu Tanaka is with the Graduate School of Science and Technology, Keio University, Kanagawa 223-8522, Japan; the Department of Applied Physics and Physico-Informatics, Keio University, Kanagawa 223-8522, Japan; the Keio University Sustainable Quantum Artificial Intelligence Center (KSQAIC), Keio University, Tokyo 108-8345, Japan; the Human Biology-Microbiome-Quantum Research Center (WPI-Bio2Q), Keio University, Tokyo 108-8345, Japan; and the Green Computing Systems Research Organization (GCS), Waseda University, Tokyo 162-0042, Japan (e-mail: shu.tanaka@appi.keio.ac.jp).
}
}

\markboth{Journal of \LaTeX\ Class Files,~Vol.~18, No.~9, September~2020}%
{How to Use the IEEEtran \LaTeX \ Templates}

\maketitle

\begin{abstract}
Ising machines are expected to solve combinatorial optimization problems faster than the existing integer programming solvers. These problems, particularly those encountered in practical situations, typically involve inequality constraints. However, owing to the hardware limitations of the current Ising machines, solving combinatorial optimization problems with inequality constraints remains challenging. The Capacitated Vehicle Routing Problem (CVRP) is a typical example of a problem with inequality constraints. The objective function of the CVRP is to minimize the total distance traveled by each vehicle while limiting the total demand of customers served by a single vehicle to the vehicle’s capacity. The CVRP is classified as $\mathcal{NP}$-hard and, thus, is commonly solved using heuristic algorithms, such as column generation. Column generation attempts to iteratively generate only the promising routes, as the number of feasible routes increases exponentially. Within this framework, the CVRP is formulated as a set cover problem. The corresponding dual solutions are used to define the pricing subproblem, which is intended to create a new route. By applying Ising machines to this pricing subproblem, the overall computation time can be reduced. This study aims to solve combinatorial optimization problems with inequality constraints using a hybrid algorithm that combines column generation and Ising machines, thereby extending the applications of the latter. We parameterize the difficulty of the inequality constraints and demonstrate that our annealing-assisted column generation can converge to a better lower bound. 
\end{abstract}


\input{introduction}
\input{ising_machine}
\input{cvrp}
\input{cg}
\input{method}
\input{results}
\input{conclusion}
\input{acknowledgments}

\bibliographystyle{IEEEtran}
\bibliography{ieee-ref}

\vspace{11pt}

\vspace{-33pt}
\begin{IEEEbiography}[{\includegraphics[width=1in,height=1.25in,clip,keepaspectratio]{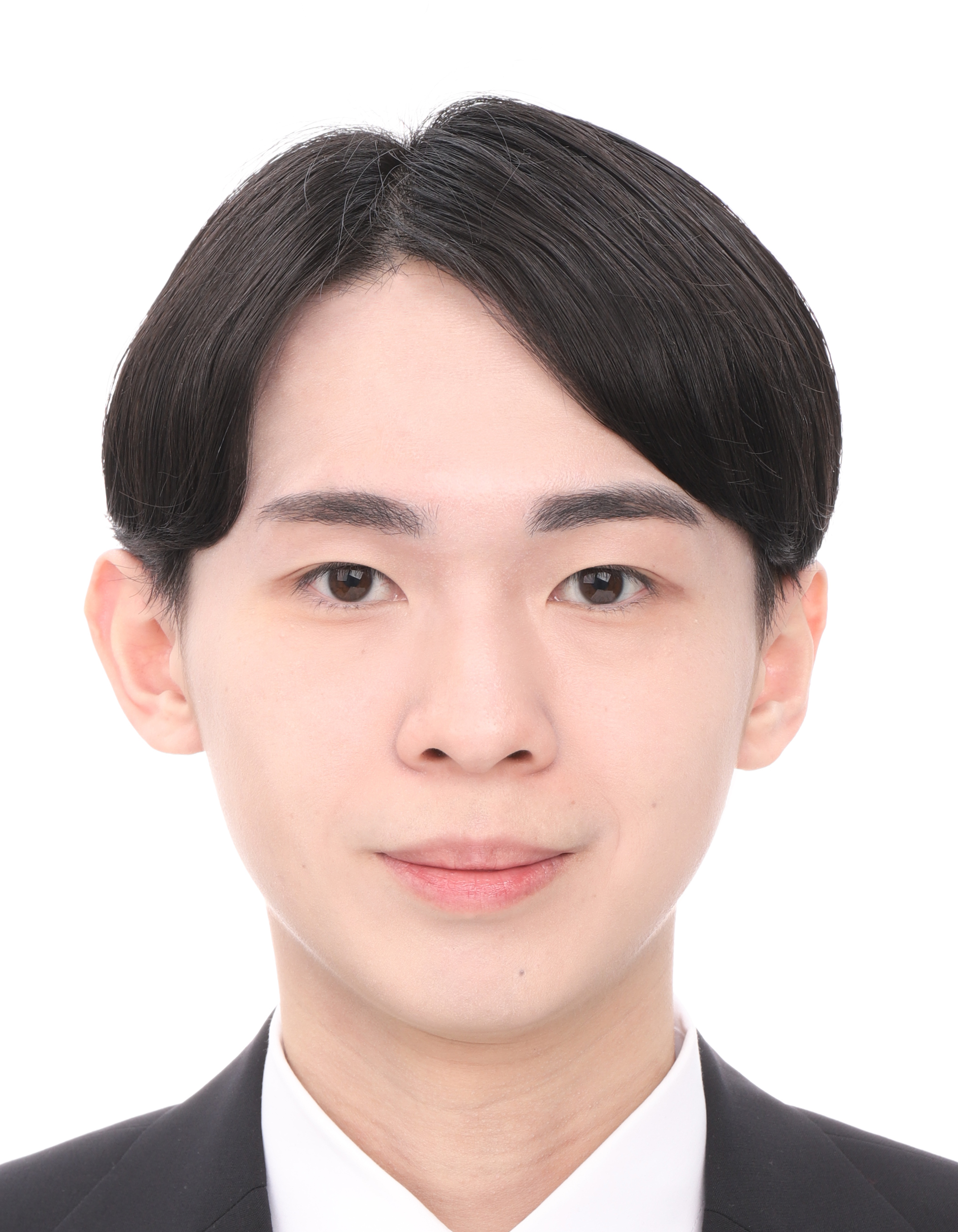}}]{Hiroshi Kanai}
received his B.~E. degree in Applied Physics and Physico-Informatics from Keio University, Kanagawa, Japan, in 2023. 
He is currently pursuing a M.~Eng. degree in Fundamental Science and Technology at the same university.
His research interests include mathematical optimization, quantum annealing, and Ising machines.
\end{IEEEbiography}

\vspace{11pt}

\vspace{-33pt}
\begin{IEEEbiography}[{\includegraphics[width=1in,height=1.5in,clip,keepaspectratio]{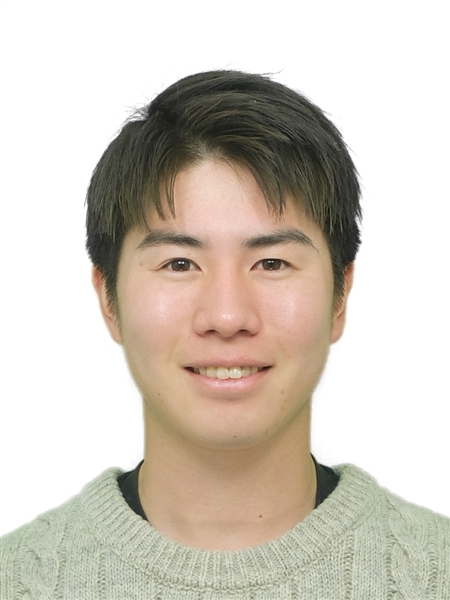}}]{Masashi Yamashita}
received his B.~E. and M.~Eng degrees in Applied Physics and Physico-Informatics from Keio University, Kanagawa, Japan, in 2022 and 2024, respectively. 
He is currently engaged in product design at Recruit Co., Ltd., Tokyo, Japan. 
His research interests include mathematical optimization, quantum annealing, Ising machines, and machine learning.
\end{IEEEbiography}

\vspace{-33pt}
\begin{IEEEbiography}[{\includegraphics[width=1in,height=1.3in,clip,keepaspectratio]{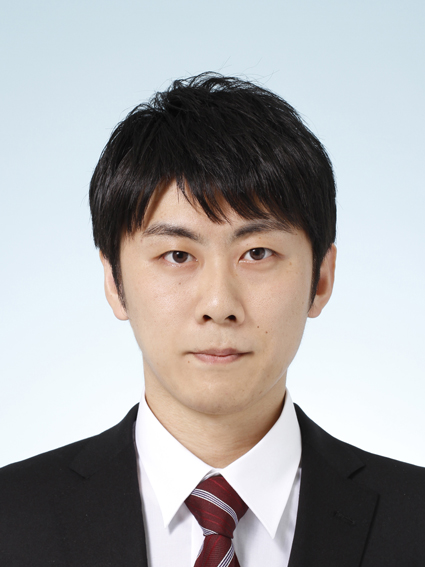}}]{Kotaro Tanahashi}
received his M.~Eng. degree from Kyoto University, Kyoto, Japan, in 2015. He has experience working as a machine learning engineer at Recruit Co., Ltd., and is working for Turing Inc., Tokyo, Japan. He also serves as a project manager for MITOU Target Program of the Information-Technology Promotion Agency (IPA).
His research interests include mathematical optimization, quantum annealing, Ising machines, machine learning, and autonomous driving.
\end{IEEEbiography}

\vspace{-33pt}
\begin{IEEEbiography}[{\includegraphics[width=1in,height=1.5in,clip,keepaspectratio]{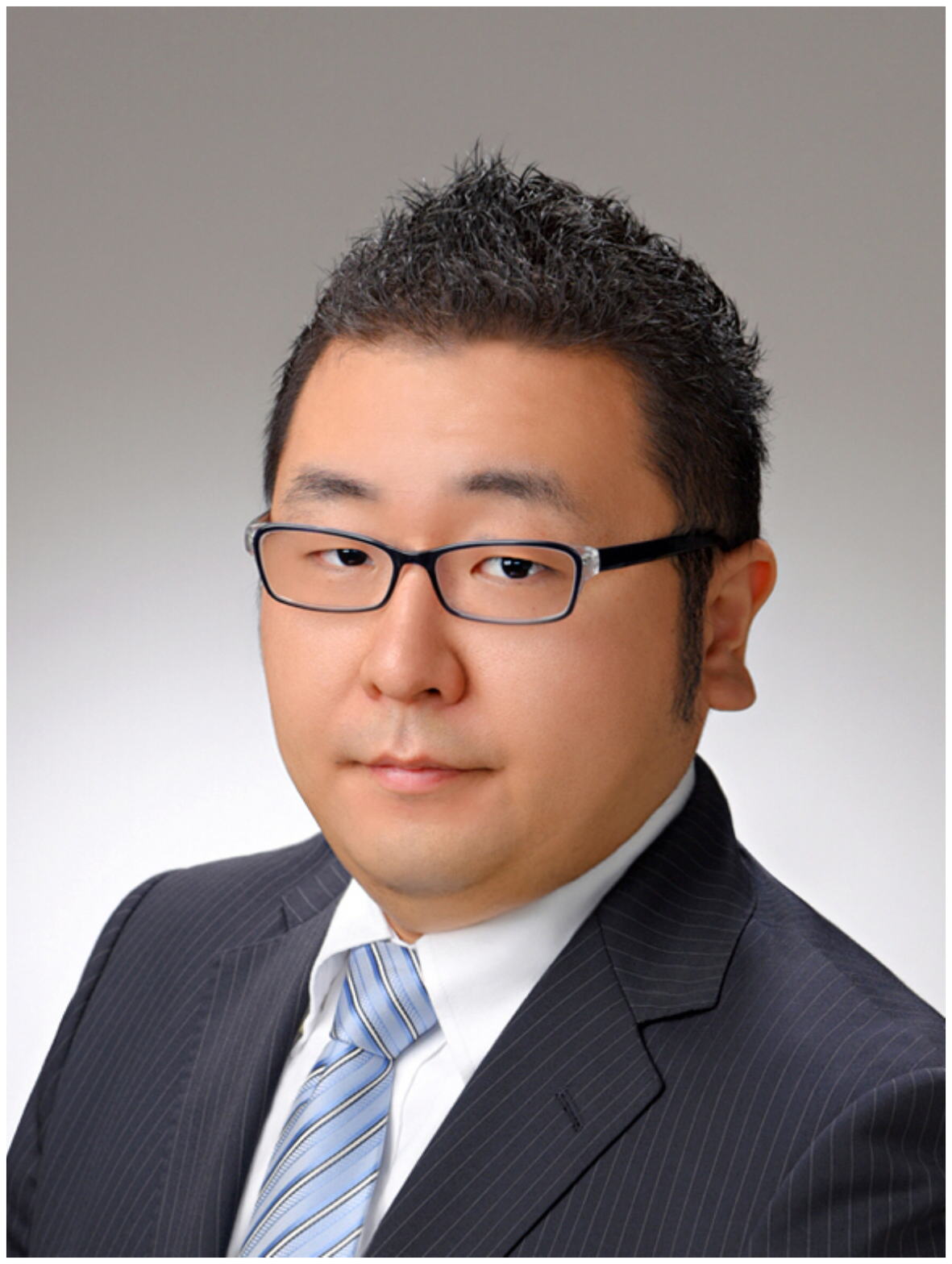}}]{Shu Tanaka} (Member, IEEE) received his B.~Sci. degree from the Tokyo Institute of Technology, Tokyo, Japan, in 2003, and his M.~Sci. and Ph.~D degrees from the University of Tokyo, Tokyo, Japan, in 2005 and 2008, respectively.
He is currently an Associate Professor in the Department of Applied Physics and Physico-Informatics, Keio University, a chair of the Keio University Sustainable Quantum Artificial Intelligence Center (KSQAIC), Keio University, and a Core Director at the Human Biology-Microbiome-Quantum Research Center (Bio2Q), Keio University. 
He is also a visiting associate professor at the Green Computing Systems Research Organization (GCS), Waseda University. 
His research interests include quantum annealing, Ising machines, quantum computing, statistical mechanics, and materials science. 
He is a member of the Physical Society of Japan (JPS), and the Information Processing Society of Japan (IPSJ).
\end{IEEEbiography}

\end{document}

%% file: introduction.tex
\section{Introduction}
\label{Sec:intro}
\IEEEPARstart{C}{ombinatorial} optimization problems entail identifying the optimal solutions that, among numerous candidate solutions, satisfy the defined constraints. 
These problems find applications in various fields, including marketing, drug development, and routing. 
If $\bm{x}$ represents a vector of the decision variables with combinatorial structure, a combinatorial optimization problems is one that seeks to obtain the solution $\bm{x}^*$ that minimizes the objective function $f(\bm{x})$, i.e.:
\begin{equation}
\label{eq:def_comb}
    \bm x^\star = \underset{\bm x} {\operatorname{argmin}} f(\bm x) \quad \bm x \in \mathcal{F},
\end{equation}
where $\mathcal{F}$ denotes the specific equality or inequality constraints of the problem. 
Ising machines are expected to provide polynomial improvements in both time complexity and scalability for the combinatorial optimization problems compared to conventional integer programming solvers~\cite{mohseni2022}.
Such enhancements are crucial, given that most combinatorial optimization problems belong to the $\mathcal{NP}$-hard class, wherein even slight variations in the coefficient of the exponential function can substantially reduce computation time.
Combinatorial optimization problems can be formulated using the Ising model, one of the most fundamental statistical mechanical models, so that the model's ground states correspond to the optimal solutions of combinatorial optimization problems~\cite{nishimori2001, lucas2014, tanaka2017, tanahashi2019}. 
However, solving combinatorial optimization problems with numerous inequality constraints poses challenges due to the current hardware limitations of Ising machines. 
These limitations are manifested in the scarcity of available parameters and topological issues to express the interactions in the corresponding Ising model~\cite{lechner2015, tomas2016, oku2020, shirai2020, kikuchi2023-1, kikuchi2023-2}. 
Furthermore, the performance of Ising machines heavily relies on the formulations of the problems~\cite{sinno2023}. 
In this study, we focus on the Capacitated Vehicle Routing Problem (CVRP) as a prototypical optimization problem with inequality constraints. 

The Vehicle Routing Problem (VRP), introduced by Dantzig and Ramser~\cite{dantzig1959}, exemplifies a multiobjective transportation network design and routing problem~\cite{john1993}. 
Its goal is to determine the shortest set of routes where each customer is visited exactly once. 
The objective function is defined as the total distance covered by each vehicle, with each vehicle required to depart from and return to its depot. 
In addition to these rudimentary premises, several VRP variants incorporate other practical constraints such as multiple depots and time windows~\cite{weise2009}. 
Of these variants, the CVRP is a particularly representative case whose properties and classical approaches have been extensively studied. 
The CVRP introduces an additional inequality constraint that the total demand of the customers visited by a single vehicle must not exceed the vehicle's capacity. 
Being $\mathcal{NP}$-hard, the CVRP is commonly handled through heuristic algorithms, such as construction methods~\cite{clarke1964, fisher1981}, column generation~\cite{dantzig1961, Gilmore1961, Gilmore1963}, and local search~\cite{paul1993}.
In particular, the column generation scheme aims to iteratively generate only the promising routes, given the exponential increase in feasible routes. 
Formulating the CVRP as a set cover problem, termed as a master problem, involves using its duals to define the pricing subproblem with capacity constraints. 
This pricing subproblem plays a critical role in creating new routes that could potentially reduce the current objective value but often becomes the bottleneck of the entire algorithm. 
Recently, Ising machines have been suggested as capable of solving this pricing subproblem more efficiently than the existing solvers. 
Relevant research in this area~\cite{ossorio2022, da2023, hirama2023} has applied hybrid algorithm combining column generation and Ising machines to solve other optimization problems.
While previous studies have been problem-oriented, our research focuses attention on the inequality constraints of certain optimization problems. 

Extended versions of branch-and-price algorithms or active set methods are commonly employed to obtain exact solutions for the CVRP. 
Specifically, the branch-and-cut-and-price method~\cite{fukasawa2006} accurately solved the problem when the number of customers was $N \simeq 100$. 
In addition, recent efforts have explored the use of Ising machines to tackle larger instances of the CVRP~\cite{feld2019,  irie2019, sinno2023, bao2024}. 
Nevertheless, the aforementioned hardware limitations prevent Ising machines from directly finding the optimal solutions for the CVRP. 
The subsequent section elaborates on these hardware limitations.  
This study aims to extend the application of Ising machines to combinatorial optimization problems with inequality constraints.

The remainder of this paper is structured as follows.
We commence by introducing Ising machines and their intricacies in Section \ref{Sec:Ising_machines}. 
In Section \ref{Sec:cvrp}, we present the conventional CVRP formulation. 
Section \ref{Sec:cg} outlines the principles of the column generation algorithm. 
The proposed methods are detailed in Section \ref{Sec:methods}. 
We present the obtained results in Section \ref{Sec:results}.
Lastly, we draw conclusions in section \ref{Sec:conclusion}.

%% file: ising_machine.tex
\section{Ising machine}
\label{Sec:Ising_machines}
\noindent As noted earlier, combinatorial optimization problems are ubiquitous in our society. 
The most well-known combinatorial optimization problems belong to the $\mathcal{NP}$-hard class. 
However, solving these $\mathcal{NP}$-hard problems is time-consuming. 
Whether $\mathcal{NP}$-hard problems can be solved in polynomial time remains to be determined.
Thus, approximation and heuristic algorithms are currently being developed for this task.
Recent years have seen the brilliant performances of Ising machines that aim to obtain the ground state of the Ising model~\cite{mohseni2022, nishimori2001, lucas2014, tanaka2017, tanahashi2019}.
Here, the state is defined as the spin configuration of the Ising model, and the lowest-energy state is specifically called the ground state.  
This Ising problem is also $\mathcal{NP}$-hard, and many of the combinatorial optimization problems can be mapped to Ising model as well as its equivalent expression, the Quadratic Unconstrained Binary Optimization (QUBO) model.
Therefore, improvements over the classical algorithms for these Ising problems can also have a significant influence on other combinatorial optimization problems as well~\cite{dutta2021}. 

The Ising model is defined on an undirected weighted graph $G=(V, E)$, where $V$ and $E$ denote the vertex and undirected-edge set, respectively. 
Here, $N = |V|$ and their elements are given by $V = \{v_0, v_1, \dotsc, v_{N-1}\}$ and $E=\{(i,j)|v_i, v_j\in V\}$.
Let $s_i$ be an Ising variable that takes either $+1$ or $-1$. 
The Ising Hamiltonian is then given by:
\begin{equation}
\begin{split}
\label{eq:Ising_model}
    & \mathcal{H}(\{s\}) = - \sum_{0 \leq i < j \leq N-1} J_{i,j}s_i s_j - \sum_{i=0}^{N-1} h_i s_i, \\
    & s_i \in \{+1, -1\} \qquad  ^\forall v_i \in V, \\ 
    & \{s\} = \{s_0, \dotsc, s_{N-1}\}.
\end{split}
\end{equation}
$J_{i, j}$ represents the real-valued interaction coefficient between $v_i$ and $v_j$. 
The real-valued local bias $h_i$ can be added to $v_i$, which is also called the local magnetic field. 
Both $J_{i, j}$ and $h_i$ can be set in advance as constants.
If $J_{i, j}$ is a nonzero value for every pair of $v_i$ and $v_j$ in the graph $G$, the graph is stated to be fully connected. 
Moreover, a subset of vertices such that each and every pair of $v_i$ and $v_j$ is adjacent is called a clique. 
Binary variables are sometimes appropriate for formulating combinatorial optimization problems.
The binary variables are denoted by $x_i\in\{0, 1\}$.
Using the following transformation rule, the Ising model can be converted to the QUBO model and vice versa. 
\begin{equation}
    x_i = \frac{s_1 + 1}{2}\qquad ^\forall v_i \in V.
\end{equation}
The corresponding objective function is expressed as follows:
\begin{equation}
\begin{split}
    & \mathcal{E}(\{x\}) = \sum_{0 \leq i \leq j \leq N-1} \mathcal{A}_{i,j}x_i x_j, \\
    & x_i \in \{0, 1\} \quad ^\forall v_i \in V, \ \{x\} = \{x_0, \dotsc, x_{N-1}\},
\end{split}
\end{equation}
where $\mathcal{A}_{i, j}$ is the off-diagonal coefficient and $\mathcal{A}_{i,i}$ is the diagonal coefficient. 
Each element $A_{i, j}$ is a real constant.
Note that we used $x_i x_i = x_i$ based on the basic property of binary variables; therefore, the equation above includes a linear term.
The penalty function method is usually a good way forward for formulating optimization problems with constraints.
In the penalty function method, the equality and inequality constraints are expressed as a penalty function such that the penalty terms take a positive value if their corresponding constraints are violated. 
The linear combination of the penalty function and the objective function is a new objective function that should be optimized. 
Because we consider the minimization problems, the violation increases the objective values and prevents us from obtaining the global optima. 

To obtain the lowest-energy states of the Ising Hamiltonian or the energy function, the various underlying operating mechanisms for Ising machines have been devised and implemented: Quantum Annealing (QA)~\cite{kadowaki1998}, Simulated Annealing (SA)~\cite{kirkpatrick1983, johnson1989}, Coherent Ising Machine (CIM)~\cite{wang2013, yamamoto2017, toshimori2021}, and Simulated Quantum Annealing (SQA)~\cite{crosson2016, okuyama2017}. 

When we make use of Ising machines to solve optimization problems, mapping from the logical model to the actual interaction topology, which is called minor embedding, is required.
We refer to the decision variables in the logical model as the logical spins, which are distinguished from the physical spins in the actual circuit. 
We will later show that logical spins in a logical model sometimes need to be expressed as multiple physical spins in real devices. 
In this study, we employed Fixstars Amplify Annealing Engine.
Fixstars Amplify Annealing Engine is a Graphics Processing Unit (GPU)-based Ising machine operated using an SA-like algorithm. 
The maximum number of available logical spins for a fully connected graph is $131,072$. 

\begin{figure}[t]
    \centering
    \includegraphics[keepaspectratio, scale=0.5]{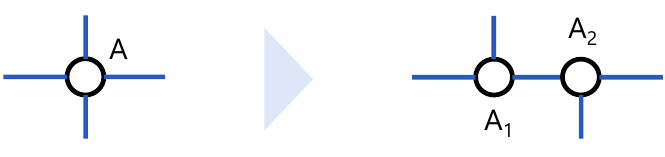}
    \caption{The interaction topology problem on real devices. Even though the logical spin is connected to the four neighboring bits in the logical model, it may not be possible to realize this if one physical spin can be connected to three physical spins. To solve this problem, multiple physical spins are used to express a single logical spin. These equivalent physical spins are strongly linked so that they can have the same value. }
    \label{fig:topology_issue}
\end{figure}

Although Ising machines are undergoing significant progress in terms of their ability to handle a wide range of problems, three fundamental hardware limitations must be considered.
First, the number of available bits is limited in real devices. 
For instance, the D-Wave Advantage~\cite{kelly2020} can handle more than $5000$ qubits as a sparse graph and $177$ logical spins as a clique. 
Therefore, it remains a challenge to solve problems larger than $N=177$ that can be defined on fully connected graphs.
Second, some of the interactions in the logical model cannot be directly expressed on real devices because of interaction topology issue~\cite{lechner2015}. 
As shown in Figure~\ref{fig:topology_issue}, although the logical spin is connected to four neighboring bits in the logical model, it may be impossible to connect them directly if the possible number of neighboring bits is three. 
Thus, multiple physical spins are used to express a single logical spin by setting the sufficiently large interactions between those equivalent physical spins. 
This interaction topology issue hinders us from solving larger problems whose corresponding undirected graphs have dense connectivity. 
Finally, the bit widths of the interactions and the local magnetic fields in the physical model are limited.  
Therefore, when mapping the logical models onto the physical models, the ground states of the original model and converted models can be different. 
To avoid this, bit-width reduction methods that maintain the correspondence between the logical model and the physical models have been studied~\cite{oku2020, kikuchi2023-2}.

Owing to these limitations, Ising machines have difficulty in solving problems whose corresponding graphs require numerous variables and dense connectivity. 
However, the actual optimization problems in our society often demand both of them.
Insofar as we encounter such complex problems, reducing the number of variables and breaking down the problem into smaller ones are indispensable. 
A typical example of these optimization problems is CVRP, which is explained in the following section. 

%% file: cvrp.tex
\section{CVRP}
\label{Sec:cvrp}
\begin{figure}[b]
    \centering
    \includegraphics[keepaspectratio, scale=0.5]{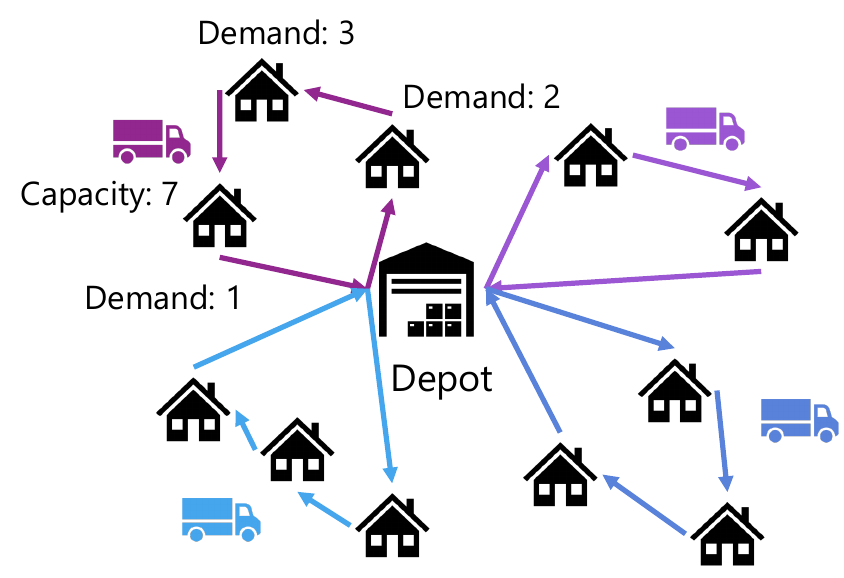}
    \caption{The brief image of CVRP. Each vehicle departs from and returns to the depot. Each customer has demand $d_i$ and the total demand of one single route must not exceed the capacity of each vehicle $Q$.}
    \label{fig:image_cvrp}
\end{figure}
\noindent CVRP (Fig.~\ref{fig:image_cvrp}) is defined on an undirected weighted complete graph $G=(V, E)$. 
Each customer $v_i$ has a demand $d_i$, and $U$ is the number of vehicles available. 
For simplicity, all the vehicles have the same capacity $Q$. 
The distances between $v_i$ and $v_j$ is denoted by $c_{i,j}$. 
Note that we enforce the symmetry condition for the distance, which is $c_{i, j} = c_{j, i}$ as the graph is undirected. 
$v_0$ corresponds to the depot and its demand $d_0 = 0$. 
Furthermore, $\delta(v_i)$ is an undirected edge set whose elements are the edges between $v_i$ and $v_j \in V \setminus \{v_i\}$ for all $j$. 
The number of movements between $v_i$ and $v_j$ is given by decision variable $x_{i,j}$. 
We also define $F(S)$ as the minimum number of vehicles required to visit all the customers in $S \subsetneq V$. 
The CVRP can then be formulated using the following equations~\cite{ralphs2003}:
\begin{align}
    & {\text{min}} && \sum_{(i, j) \in E}{c_{i, j} x_{i,j}}  \label{eq:cvrp_obj} \\
    &\text{s.t.} && \sum_{(0, j) \in \delta(v_0)}{x_{0, j}} = 2U, \hspace{3cm} \label{eq:cvrp_c1} \\
    & && \sum_{(i, j)\in \delta(v_i)}x_{i,j} = 2 \hspace{1.6cm} ^\forall v_i \in V \setminus \{v_0\}, \label{eq:cvrp_c2} \\
    & && \sum_{\substack{(i,j)\in E \\ i\in S, j \notin S}} x_{i,j} \geq 2 F(S) \hspace{0.7cm} ^\forall S \subset V, \quad |S| > 1, \label{eq:cvrp_c3} \\
    & && x_{i,j} \in \{0, 1\} \hspace{1.4cm} ^\forall (i,j)\in E, \quad  i, j\neq 0,\label{eq:cvrp_c4} \\
    & && x_{0,j} \in \{0, 1, 2\} \hspace{2.0cm} ^\forall v_j \in V\setminus \{v_0\}. \label{eq:cvrp_c5}
\end{align}
Constraint \eqref{eq:cvrp_c4} forces $x_{i, j}$ to be $1$ if the vehicle moves from $v_i$ to $v_j$ and $0$ otherwise. 
Note that the constraint \eqref{eq:cvrp_c4} excludes the case in which $i = 0$ or $j = 0$. 
Then, constraint \eqref{eq:cvrp_c5} complements the case where either $i$ or $j$ is $0$. 
Note also that $x_{0, j}$ can be $2$ in constraint \eqref{eq:cvrp_c5} because there might be a piston route, where a vehicle visits just one single customer and then returns to the depot. 
Constraint \eqref{eq:cvrp_c1} ensures that the exact number of vehicles is operated to deliver to each customer. 
As every vehicle must depart from and return to the depot, the sum of $x_{0, j}$ should be twice the number of vehicles $U$. 
Constraint \eqref{eq:cvrp_c2} ensures that each and every customer is visited strictly once. 
Constraint \eqref{eq:cvrp_c3} is called the rounded capacity inequality and imposes both the capacity and the subtour elimination constraints. 
By definition, Obtaining $F(S)$ is realized by solving the corresponding bin packing problem. 
Nonetheless, the bin packing problem is $\mathcal{NP}$-hard; thus, the following lower bound is commonly substituted for $F(S)$~\cite{ralphs2003, ibrahima2017}:
\begin{equation}
\label{eq:F(S)}
    F(S) = \left\lceil \frac{\sum_{v_i\in S}d_i}{Q}\right\rceil \hspace{0.7cm} ^\forall S \subset V, \quad |S| > 1.
\end{equation}
The symbol $\lceil \cdot \rceil$ denotes the ceiling function.
Obviously, this approximate version of $F(S)$ represents the average number of vehicles required to visit each customer in $S$.

CVRP can also be formulated in a different fashion, such that the decision variables are binary. 
In the natural course of events, a naive QUBO model for CVRP was unveiled, which is discussed in a subsequent section.  
However, as previously stated, the use of Ising machines to solve larger CVRP does not seem appropriate because of hardware limitations.
Hence, the problem is decomposed into smaller problems to enable the use of Ising machines in larger CVRP. 
The subsequent section presents the underlying principles of column generation. 

%% file: cg.tex
\section{Column Generation}
\label{Sec:cg}
\noindent This section focuses on how the column generation scheme is applied to the CVRP. 
To begin with, we describe the set cover formulation for the CVRP. 
Thereafter, we introduce the corresponding dual problem. 
Finally, we discuss the fundamental principles of column generation.

\subsection{Set Cover Problem}
\label{subSec:MP}
We define $\Omega$ as the route set that includes all feasible routes in the specific CVRP instance. 
$\omega_k$ denotes the total distance of the $k$th route $r_k$. 
Let $A=[a_{i,k}]_{N\times|\Omega|}$ be a matrix whose element $a_{i, k}$ is $1$ if route $r_k$ contains the customer $v_i$ and $0$ if route $r_k$ does not contain the customer $v_i$. 
The set cover problem for the CVRP can then be represented by the following equation:
\begin{align}
& {\text{min}} &&\sum_{r_k \in \Omega}{\omega_{k} \theta_k} \label{eq:mp_obj}\\
&\text{s.t.} && \sum_{r_k \in \Omega}{a_{i,k}\theta_k \geq 1} &&&({}^{\forall}{v_i}\in V \setminus \{v_0\}), \label{eq:mp_c1}\\
& && \sum_{r_k \in \Omega}\theta_k = U, \label{eq:mp_c2}\\
& && \theta_k \in \mathbb{N}. \label{eq:mp_c3}
\end{align}
$\theta_k$ is the decision variable that takes $1$ if route $r_k\in \Omega$ is chosen and $0$ if $r_k\in \Omega$ is not selected for the solution. 
The optimization problem \eqref{eq:mp_obj}--\eqref{eq:mp_c3} is referred to as the master problem. 
Constraint \eqref{eq:mp_c1} guarantees that every customer is visited at least once. 
Constraint \eqref{eq:mp_c1} can be formulated as an equality constraint, but this typically slows down the convergence of the column generation. 
Consequently, an inequality form is typically used for this purpose, particularly during column generation. 
If the constraint \eqref{eq:mp_c1} is expressed as an inequality, this problem is called the a set partition problem.
The number of vehicles operated is determined by the constraint $\eqref{eq:mp_c2}$. 
Finally, the constraint \eqref{eq:mp_c3} requires the decision variables to be integers. 
This is because the additional inequality constraint $\theta_k \leq 1$ can be eliminated in a restricted master problem, that underwent linear programming relaxation, as discussed later. 
Because the optimal $\theta_k$ is $0$ or $1$ in this model, $\theta_k \geq 2$ is not optimal.

The set cover problem is known to be $\mathcal{NP}$-complete, and the greedy algorithm is a well-known approximation algorithm for this problem. 
The number of feasible routes in $\Omega$ increases exponentially, and it is virtually impossible to consider all the feasible routes. 
Therefore, column generation focuses on a subset of $\Omega$, that comprises only promising routes. 
Let $\Omega^\text{RST}$ be such a restricted route set. 
Converting $\Omega$ in the set cover problem \eqref{eq:mp_obj}--\eqref{eq:mp_c3} into $\Omega^\text{RST}$ yields the following restricted master problem:
\begin{align}
& {\text{min}} &&\sum_{r_k \in \Omega^{\text{RST}}}{\omega_{k} \theta_k} \label{eq:rmp_obj}\\
&\text{s.t.} && \sum_{r_k \in \Omega^{\text{RST}}}{a_{i,k}\theta_k \geq 1} &&&{}^{\forall}{v_i}\in V \setminus \{v_0\}, \label{eq:rmp_c1}\\
& && \sum_{r_k \in \Omega^{\text{RST}}}\theta_k = U, \label{eq:rmp_c2}\\
& && \theta_k \geq 0. \label{eq:rmp_c3}
\end{align}
The restricted master problem \eqref{eq:rmp_obj}--\eqref{eq:rmp_c3} is differentiated from the original set cover problem \eqref{eq:mp_obj}--\eqref{eq:mp_c3} because the integer variable $\theta_k$ is relaxed to a real number. 
During this procedure, we aim to optimize the restricted master problem. 
This relaxation is effective because the restricted master problem provides a tight lower bound for the original set cover problem. 
More precisely, the optimal objective value $E^\text{LP}$ of the restricted master problem asymptotically reaches the optimal objective value $E^\star$ of the original set cover problem when the number of customers $N$ approaches to infinity~\cite{desrochers1992, levi2014}.
\begin{equation}
\label{eq:theorem_tightbound}
    \lim_{N\rightarrow\infty}{\frac{1}{N}E^\text{LP}} = \lim_{N\rightarrow\infty}{\frac{1}{N}E^\star}.
\end{equation}
The following subsection introduces the dual problem of the restricted master problem, which is crucial to determine the existence of potential routes. 
\subsection{Dual Problem}
The restricted master problem \eqref{eq:mp_obj}--\eqref{eq:mp_c3} can be rewritten in the following standard inequality form:
\begin{align}
& {\text{min}} &&\sum_{r_k \in \Omega^{\text{RST}}}{\omega_{k} \theta_k} \label{eq:rmp_to_dual_obj}\\
&\text{s.t.} && \sum_{r_k \in \Omega^{\text{RST}}}{a_{i,k}\theta_k \geq 1} &&& {}^{\forall}{v_i}\in V \setminus \{v_0\}, \label{eq:rmp_to_dual_c1}\\
& && \sum_{r_k \in \Omega^{\text{RST}}}\theta_k \geq U, \label{eq:rmp_to_dual_c2}\\
& && - \sum_{r_k \in \Omega^{\text{RST}}}\theta_k \geq - U. \label{eq:rmp_to_dual_c3}
\end{align}
Multiplying both sides of constraints \eqref{eq:rmp_to_dual_c1}--\eqref{eq:rmp_to_dual_c3} by the nonnegative dual variables $y_i$, $y_0$, $y_0'$, and then adding the resulting terms, we obtain: 
\begin{equation}
    \begin{split}
    \label{eq:fml_trns}
        \sum_{r_k \in\Omega^\text{RST}}{\left[\sum_{v_i\in V \setminus\{v_0\}}a_{i,k}y_i + (y_0 - y_0')\right]}\theta_k \hspace{1cm}\\
        \geq \sum_{v_i \in V \setminus\{v_0\}}{y_i} + (y_0 - y_0')U.
    \end{split}
\end{equation}
Substituting $(y_0 - y_0')$ with $y_0 \in \mathbb{R}$, the relationship above becomes equivalent to
\begin{equation}
\label{eq:redefinition_y0}
    \sum_{r_k \in\Omega^\text{RST}}{\left(\sum_{v_i\in V \setminus\{v_0\}}a_{i,k}y_i + y_0\right)}\theta_k \geq \sum_{v_i \in V \setminus\{v_0\}}{y_i} + y_0 U.
\end{equation}
Assuming the coefficients on the left side of \eqref{eq:redefinition_y0} are smaller than $\omega_k$, 
\begin{equation}
    \omega_{k} \geq \sum_{v_i\in V \setminus\{v_0\}}a_{i,k}y_i + y_0 \qquad {}^\forall r_k \in \Omega^\text{RST},
\end{equation}
we obtain the following relationship:
\begin{equation}
\label{eq:lowerbound_mp}
    \begin{split}
        \sum_{r_k \in\Omega^\text{RST}}{\omega_{k} \theta_k} \geq \sum_{r_k \in\Omega^\text{RST}}{\left(\sum_{v_i\in V \setminus\{v_0\}}a_{i,k}y_i + y_0\right)}\theta_k \hspace{0.8cm}\\ 
        \geq \sum_{v_i \in V \setminus\{v_0\}}{y_i} + y_0 U. \hspace{1.0cm}
    \end{split}
\end{equation}
According to the inequality \eqref{eq:lowerbound_mp}, the lower bound of the objective value of the restricted master problem is given by the following term:
\begin{equation}
    \sum_{v_i \in V \setminus\{v_0\}}{y_i} + y_0 U.
\end{equation}
Therefore, the dual problem of the restricted master problem \eqref{eq:rmp_obj}--\eqref{eq:rmp_c3} is
\begin{align}
& {\text{max}} &&\sum_{v_i \in V \setminus\{v_0\}}{y_i} + y_0 U \label{eq:dp_obj}\\
&\text{s.t.} &&  \sum_{v_i\in V \setminus\{v_0\}}a_{i,k}y_i + y_0 \leq \omega_{k} \qquad{}^\forall r_k \in \Omega^\text{RST}, \label{eq:dp_c1}\\
& && y_i \geq 0 \hspace{3.23cm} {}^{\forall}{v_i}\in V \setminus \{v_0\}, \label{eq:dp_c2}\\
& && y_0 \in \mathbb{R}. \label{eq:dp_c3}
\end{align}
The symbol $y_i ({}^{\forall}{v_i}\in V \setminus \{v_0\})$ denotes a nonnegative dual variable corresponding to each customer $v_i$.
Variable $y_0$ is a real dual variable corresponding to the depot. 

\subsection{The Principle of Column Generation}
Prior to introducing the pricing subproblem, we describe the fundamental principles underlying the column generation scheme. 
The feasible region in the restricted master problem \eqref{eq:rmp_obj}--\eqref{eq:rmp_c3} expands proportionally with the number of generated columns. 
Conversely, the feasible region in the dual problem \eqref{eq:dp_obj}--\eqref{eq:dp_c3} contracts in proportion to the increase in the number of potential routes generated.
This fact is of great importance because it implies that the dual solution $\bm y_1^\star$, which is optimal in the feasible region $\Omega_1^\text{RST}$, is a global optimum in $\Omega$ if $\bm y_1^\star$ is feasible in $\Omega$~\cite{feillet2010}.
Consequently, if the global optimum has not been determined from the current restricted master problem, there should be a concealed new route $r_k^\text{new}$ whose inequality constraints are violated by the current dual solution.
This new route can be identified by solving the following pricing subproblem, which reflects the violation of the inequality constraints in the dual problem:
\begin{align}
    & {\text{min}} && \omega_{k} - \sum_{v_i\in V \setminus\{v_0\}}a_{i,k}y_i^\star - y_0^\star \label{eq:sp_obj}\\
    &\text{s.t.} &&  \sum_{v_i\in V \setminus\{v_0\}}a_{i,k}d_i \leq Q. \label{eq:sp_c1}
\end{align}
\begin{figure}[b]
    \centering
    \includegraphics[keepaspectratio, scale=0.5]{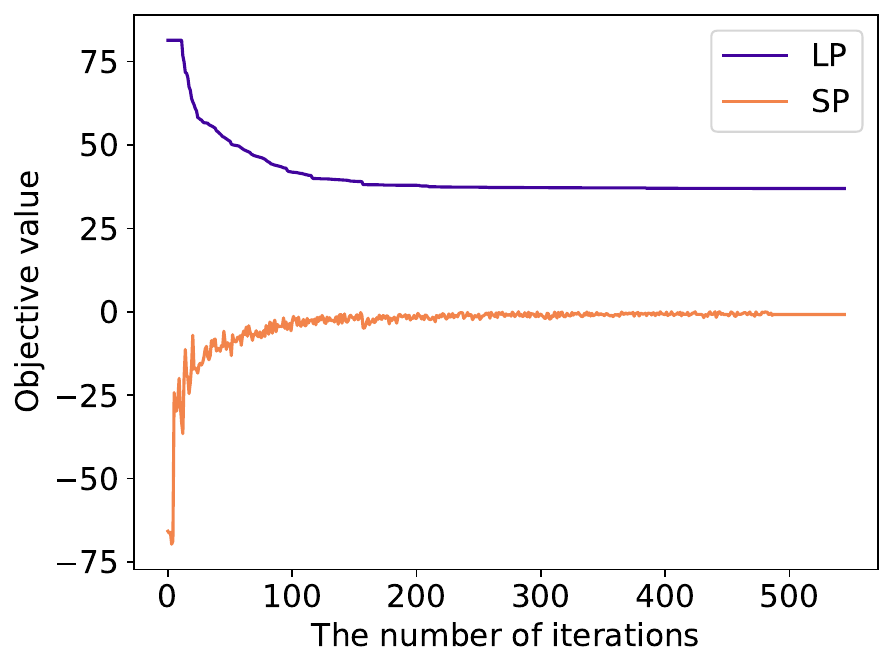}
    \caption{Time evolution of the objective value of the restricted master problem and the pricing subproblem. The purple line (LP) represents the objective value for the restricted master problem, while the orange line (SP) depicts the reduced cost of the pricing subproblem.}
    \label{fig:example_cg_obj}
\end{figure}
The objective function \eqref{eq:sp_obj} is called the reduced cost $R_k$. 
The prior discussion asserts that route $r_k$ with a negative reduced cost, i.e., $R_k < 0$ has the potential to lower the current objective value. 
When $R_k < 0$, the iterative process of solving the restricted master problem, the dual problem, and the pricing subproblem continues. 
Figure~\ref{fig:example_cg_obj} illustrates the evolution of the relevant objective values throughout the column generation process.  
Initially, a random set of routes is generated, the total distance of which is typically far from the optimal solution. 
It can be observed that the corresponding reduced cost is negative at the initial iteration. 
As the iteration progresses, the objective value of the restricted master problem improves, whereas the reduced cost approaches zero. 
If a new route is not acquired and the reduced cost is negative, the algorithm is terminated. 
After this iterative process, we solved the original set cover problem to obtain the optimal routes. 
However, these tentatively good routes may still be problematic owing to the set cover formulation; that is, some of the chosen routes visit the same customers. 
To resolve this undesirable situation, we finally solve the set partition problem, where the inequality constraint \eqref{eq:mp_c1} is modified to an equality constraint.
Note that the dual solutions $y_i^\star$ in the pricing subproblem determine the order of priority for each customer.
This is because customers whose dual-solution value is high can lower their reduced cost if the route contains such customers.

\begin{figure}[t]
    \centering
    \includegraphics[keepaspectratio, scale=0.3]{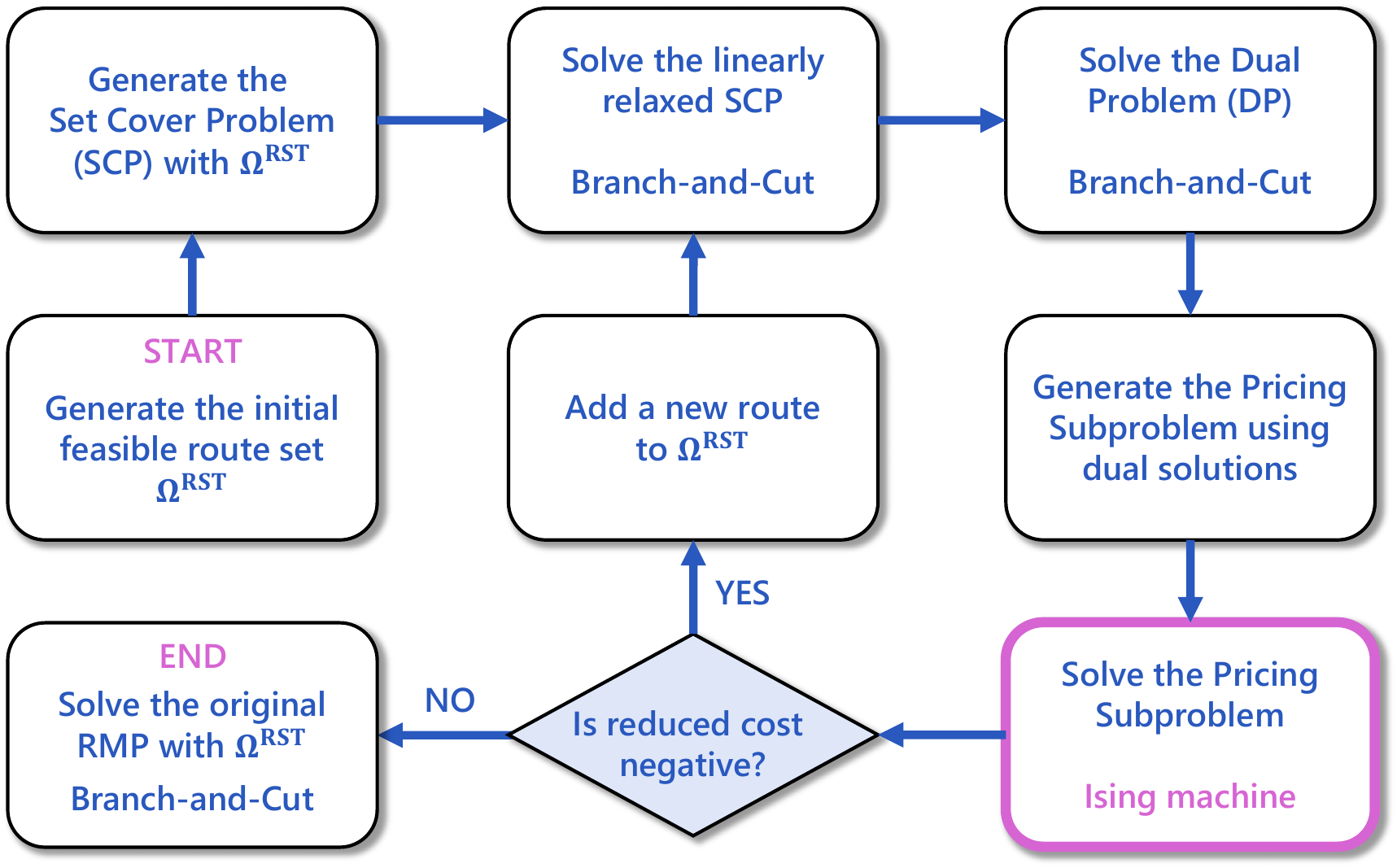}
    \caption{A flowchart of column generation. To optimize the set cover problem and the dual problem, we used CBC solver of PuLP. Besides, we applied an Ising machine to the pricing subproblem to speed up and handle this bottleneck.}
    \label{fig:flow_chart}
\end{figure}

A new route $r_k^\text{new}\in\Omega\setminus\Omega^\text{RST}$ is obtained as an output equivalent to the column of the matrix $A = [a_{i,k}]_{N\times|\Omega|}$. 
Therefore, the sequence of optimization procedures described here is called column generation.
Although the pricing subproblem is the shortest path problem with negative weight edges, it entails the capacity constraints \eqref{eq:sp_c1} and is known to be $\mathcal{NP}$-hard.
One solution to this difficulty is to allow each customer to be visited more than once by modifying the matrix element $a_{i, k}$ alleviates this difficulty. 
Consequently, this modification enables the execution of the column generation in pseudopolynomial time using dynamic programming~\cite{levi2014}. 
With respect to the restricted master problem and the dual problem, we used the COIN-OR branch and cut (CBC) solver offered by an open-source library PuLP to optimize them.
An overall flowchart of the column generation method is shown in Figure~\ref{fig:flow_chart}.

In this study, we utilized a GPU-based Ising machine to solve this pricing subproblem, without modifying the matrix element $a_{i,k}$.
The next section is devoted to explaining how the pricing subproblem is solved using an Ising machine. 

%% file: method.tex
\section{Methods}
\label{Sec:methods}
\noindent In this section, we describe how our annealing-assisted column generation is implemented. 
As explained earlier, to solve the combinatorial optimization problem, the objective function must be written in either QUBO or the Ising model. 
The first subsection describes the formulation of the pricing subproblem as a QUBO model.
In the second subsection, we explain the solution of the CVRP using only an Ising machine.
The preprocessing and postprocessing techniques related to neighborhood searches are presented in the following subsections.
Finally, we propose a method in which the generated columns have low similarity to each other, which is realized by fixing binary variables. 

\subsection{Formulation of the Pricing Subproblem as a QUBO Model}
The previous section introduced the pricing subproblem expressed in the Equations ~\eqref{eq:sp_obj} and \eqref{eq:sp_c1}. 
Let $q_{t, i}$ be a binary variable that takes $1$ if the vehicle visits the customer $v_i$ at the $t$th step, and $0$ otherwise.
Using this binary variable $q_{t, i}$, the QUBO model of the pricing subproblem is as follows:
\begin{align}
    \mathcal{H}_\text{sp} & = \mathcal{H}_{\text{dist}} + \mathcal{H}_\text{price} + p\left(\mathcal{H}_\text{c1} + \mathcal{H}_\text{c2} + \mathcal{H}_\text{c3} \right), \label{eq:sp_qubo_obj}\\
    \mathcal{H}_\text{dist} & = \sum_{t=1}^{T-1} \sum_{i=0}^{N-1} \sum_{j=0}^{N-1} c_{i,j} q_{t, i} q_{t+1, j}, \label{eq:sp_qubo_dist} \\
    \mathcal{H}_\text{price} & = - \sum_{i=1}^{N-1} {y_i^\star}\sum_{t=1}^T {q_{t, i} - y_0^\star }, \label{eq:sp_qubo_price}\\
    \mathcal{H}_\text{c1} & = \sum_{i=1}^{N-1} {\left( \sum_{t=1}^T q_{t, i} - z_i \right)^2}, \label{eq:sp_qubo_c1} \\
    \mathcal{H}_\text{c2} & = \sum_{t=1}^T {\left( \sum_{i=0}^{N-1} q_{t, i} - 1 \right)^2}, \label{eq:sp_qubo_c2}\\ 
    \mathcal{H}_\text{c3} & = \left( \sum_{i=1}^{N-1} \sum_{t=1}^{T} {d_{i} q_{t, i} - h}\right)^2, \label{eq:sp_qubo_c3} \\
    q_{t, i} & \in \{0, 1\},
\end{align}
where $p$ is a penalty coefficient and defined as $p:=\max_{i\neq j}{[c_{i,j}-y^\star_i]}$ to ensure that the reduced cost term is of the same order of magnitude as for the penalty term.
$T$ is the maximum number of steps that each vehicle can take and is determined to satisfy $T \geq N / U$. 
$z_i \in \{0, 1\}$ in Eq. \eqref{eq:sp_qubo_c1} is a slack variable.
Equations \eqref{eq:sp_qubo_dist} and \eqref{eq:sp_qubo_price} correspond to the objective function of the pricing subproblem.
The terms $\mathcal{H}_\text{c1}$, $\mathcal{H}_\text{c2}$, and $\mathcal{H}_\text{c3}$ are equivalent to the penalty terms. 
$\mathcal{H}_\text{c1}$ stipulates that each customer should be visited at most once.
At each time step $t$, the vehicle is assumed to visit only a single customer because of the term $\mathcal{H}_\text{c2}$. 
Because only one variable with a value of $1$ exists in the same row, these types of constraints are referred to as the one-hot constraints.
The capacity constraints of CVRP were converted into equality constraints to adapt the penalty function method in $\mathcal{H}_\text{c3}$.
Here, $h$ in \eqref{eq:sp_qubo_c3} is a polynomial used to implement the conversion. 
Fixstars Amplify Annealing Engine selects $h$ with the least number of slack variables from the following three options:
\begin{align}
    h & = 1 + \sum_{k=1}^{Q-1} {h_k}, \label{eq:unary_h} \\
    h & = 1 + \sum_{k=1}^{\lfloor\log_2{Q}\rfloor}{2^k h_k} + \cdots, \label{eq:binary_h} \\
    h & = 1 + \sum_{k=1}^{\lfloor\sqrt{Q}\rfloor}kh_k + \sum_{k=1}^{\lfloor\sqrt{Q}\rfloor - 1}{k h_{k+\lfloor\sqrt{Q}\rfloor}} + \cdots, \label{eq:linear_h}
\end{align}
where $h_k\in\{0, 1\}$ denotes a slack variable. 
All these polynomials are used to encode integers, that is, multiple discrete values, using binary variables. 
Hence, these encoding methods are known as binary-integer encoding~\cite{tamura2021, chen2021}.

\begin{figure}[t]
    \centering
    \includegraphics[keepaspectratio, scale=0.5]{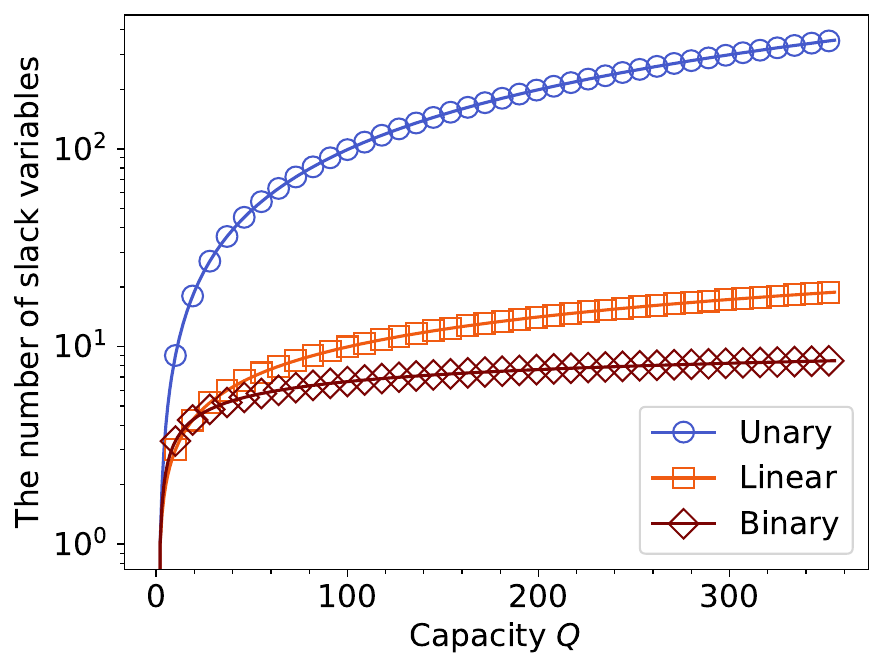}
    \caption{The number of slack variables needed to use three respective encoding methods: unary encoding, binary encoding, and linear encoding.}
    \label{fig:num_slack_variables}
\end{figure}

The first polynomial \eqref{eq:unary_h} is referred to as unary encoding, and can represent integers through the summation of slack variables that take $1$. 
The number of slack variables required for this encoding method is equal to $(Q-1)$.
Second, binary encoding, expressed as a polynomial \eqref{eq:binary_h}, employs binary notation to encode integers.
In contrast to unary encoding, binary encoding may include superfluous integers that have a detrimental effect on the inequality constraints.
To circumvent this issue, the constant terms are subtracted from the polynomial \eqref{eq:binary_h}. 
The number of slack variables required for binary encoding is estimated as $\mathcal{O}(\log{Q})$.
Finally, the polynomial \eqref{eq:linear_h} is called linear encoding. 
Linear encoding utilizes a linear combination of slack variables.
The number of slack variables required for the linear encoding is reduced to $\mathcal{O}(\sqrt{Q - 1})$. 
Figure~\ref{fig:num_slack_variables} shows the number of slack variables required for the three binary-integer encoding techniques.  
As shown later, we prepared five different problem settings with capacities of $26$, $60$, $122$, $169$, and $355$. 
Consequently, binary encoding is typically selected for the numerical simulations because it is likely to require the fewest number of slack variables. 
\begin{table}[t]
    \caption{QUBO table for $q_{t, i}$.}
    \label{table:method_qubo_table}
    \begin{center}
        \begin{tabular}{c||ccccccccccc} 
        \hline
            $t$ & $v_0$ & $v_1$ & $v_2$ & $v_3$ & $v_4$ & $v_5$ & $v_6$ & $v_7$ & $v_8$ & $v_9$ & $v_{10}$ \\
            \hline \hline
            $1$ & $0$ & $0$ & $1$ & $0$ & $0$ & $0$ & $0$ & $0$ & $0$ & $0$ & $0$ \\ \hline
            $2$ & $0$ & $0$ & $0$ & $1$ & $0$ & $0$ & $0$ & $0$ & $0$ & $0$ & $0$ \\ \hline
            $3$ & $1$ & $0$ & $0$ & $0$ & $0$ & $0$ & $0$ & $0$ & $0$ & $0$ & $0$ \\ \hline
            $4$ & $0$ & $0$ & $0$ & $0$ & $1$ & $0$ & $0$ & $0$ & $0$ & $0$ & $0$\\
            \hline
            $5$ & $0$ & $0$ & $0$ & $0$ & $0$ & $0$ & $0$ & $1$ & $0$ & $0$ & $0$\\
            \hline
            $6$ & $0$ & $0$ & $0$ & $0$ & $0$ & $0$ & $1$ & $0$ & $0$ & $0$ & $0$\\
            \hline
        \end{tabular}
    \end{center}
\end{table}

\begin{figure*}[t]
\centering
\subfloat[]{\includegraphics[width=2.0in]{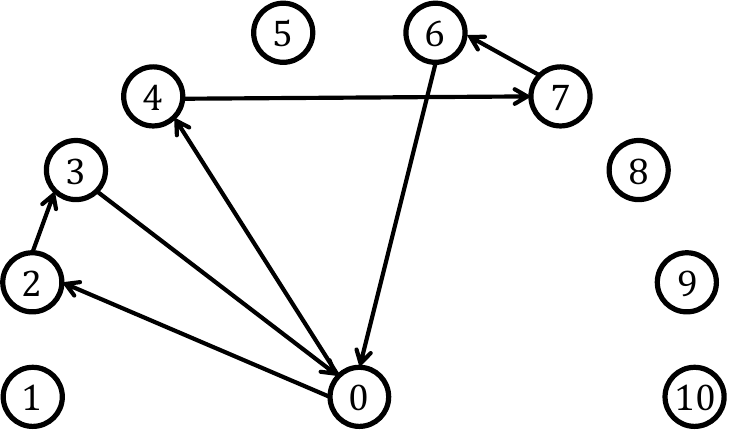}%
\label{fig:post_b}}
\hfil
\subfloat[]{\includegraphics[width=2.0in]{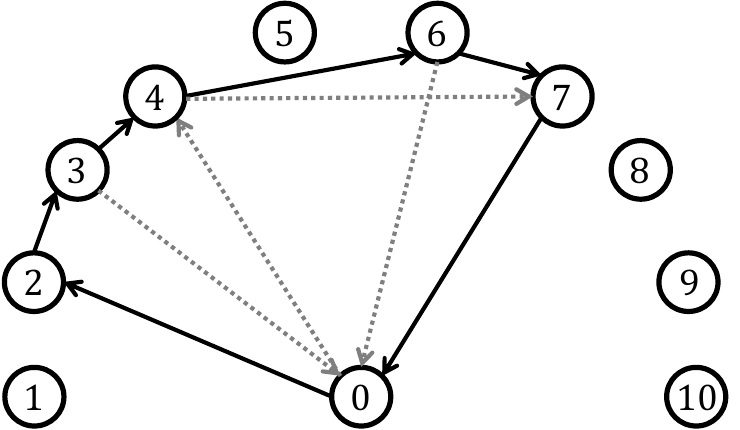}}%
\label{fig:post_a}
\caption{An example route that could be obtained by solving the pricing subproblem. (a) The initial route before the postprocessing modification. (b) The final route after the postprocessing modification.}
\label{fig:post_processing_routes}
\end{figure*}

Table~\ref{table:method_qubo_table} illustrates an example solution when the pricing subproblem is solved. 
Each row represents the $t$th vehicle movement, and each column corresponds to a depot or the customer.
Note that this QUBO table satisfies the aforementioned one-hot constraints.
In this case, the route is $[v_0 \rightarrow v_2 \rightarrow v_3 \rightarrow v_0 \rightarrow v_4 \rightarrow v_7 \rightarrow v_6 \rightarrow v_0]$. 
This route is illustrated in Figure~\ref{fig:post_processing_routes}(a).

The present objective is to minimize the total distance traveled by each vehicle; thus, we require routes to comply with the following two rules:
\begin{enumerate}
    \item The distance matrix $[c_{i, j}]_{N\times N}$ satisfies the triangle inequality:  
    \begin{equation}
        c_{i, j} + c_{j, k} \geq c_{i, k} \quad ^\forall v_i, v_j, v_k \in V.
    \end{equation}
    \item Once the vehicle returns to the depot $v_0$, it cannot depart again.
\end{enumerate}
With regard to these two rules, we have incorporated two postprocessing algorithms into the column generation: the $2$-opt neighborhood search and the triangular modification, which we will discuss in greater detail later. 
In this study, we refer to this annealing-assisted column generation as CG. 

\subsection{Firsthand QUBO Modeling of CVRP}
For comparison, we also applied Ising machines to CVRP naively. 
The general QUBO model for CVRP is relatively straightforward because we only need to extend the QUBO model \eqref{eq:sp_qubo_obj}--\eqref{eq:sp_qubo_c3} so that it can handle the multiple vehicles.  
To achieve this, the decision variable $q_{t, i}$ must be modified to $q_{t, i}^{(u)}$. 
Appended index $u$ denotes the $u$th vehicle. 
Using this newly created decision variable, we obtain:
\begin{align}
    \mathcal{H} & = \mathcal{H}_{\text{dist}} + p\left(\mathcal{H}_\text{c1} + \mathcal{H}_\text{c2} + \mathcal{H}_\text{c3} \right), \label{eq:all_qubo_obj} \\
    \mathcal{H}_\text{dist} & = \sum_{u=1}^{U}\sum_{t=1}^{T-1} \sum_{i=0}^{N-1} \sum_{j=0}^{N-1} c_{i,j} q_{t, i}^{(u)} q_{t+1, j}^{(u)}, \label{eq:all_qubo_dist} \\
    \mathcal{H}_\text{c1} & = \sum_{i=1}^{N-1} {\left( \sum_{u=1}^{U}\sum_{t=1}^T q_{t, i}^{(u)} - 1 \right)^2}, \label{eq:all_qubo_c1} \\
    \mathcal{H}_\text{c2} & = \sum_{u=1}^{U}\sum_{t=1}^T {\left( \sum_{i=0}^{N-1} q_{t, i}^{(u)} - 1 \right)^2}, \label{eq:all_qubo_c2}\\ 
    \mathcal{H}_\text{c3} & = \sum_{u=1}^{U}\left( \sum_{j=1}^{N-1} \sum_{t=1}^{T} {d_{j} q_{t, j}^{(u)} - h}\right)^2, \label{eq:all_qubo_c3} \\
    q_{t, i}^{(u)} & \in \{0,1\},
\end{align}
where $p$ is the penalty coefficient defined as $p = \max_{i\neq j}{[c_{i,j}]}$. 
The forms of the objective function and constraints are nearly identical to those of the QUBO model written for the pricing subproblem.
However, note that $\mathcal{H}_\text{c1}$ does not require slack variables because this QUBO model considers the routes of all the vehicles simultaneously. 
Hereafter in this paper, we refer to this method as AE. 
\subsection{$2$-Opt Neighborhood Search}
We adopted the $2$-opt neighborhood search for the obtained route in order to take advantage of the rule $1)$.
As the distance matrix satisfies the triangle inequality, reordering some crossing routes results in a shorter total distance for that route.  

Algorithm~\ref{alg:2-opt} presents the pseudocode for the $2$-opt neighborhood search. 
The procedure for the $2$-opt neighborhood search is as follows:
First, we select two edges, $edge_1 = (i, j)$ and $edge_2=(k, l)$ from the obtained route. 
Subsequently, the selected edges are cut, the vertices are swapped, and new edges are created, namely $edge_3 = (i, k)$ and $edge_4 = (j, l)$. 
If the lengths of the newly generated edges $edge_3$ and $edge_4$ are shorter than those of the former edge, $edge_1$ and $edge_2$, we reconnect $edge_3$ and $edge_4$ to the route. 
This sequential process constitutes one iteration of the $2$-opt neighborhood search.
This process was repeated until no further improvement was observed. 
The time complexity of an iteration of this neighborhood search is $\mathcal{O}(T^2)$. 

The $2$-opt neighborhood search was applied to CG when the initial set of feasible routes was generated and each time the pricing subproblems were solved. 
On the other hand, we applied this local search method to AE when the final solution was obtained. 
\begin{figure}[t]
\begin{algorithm}[H]
    \caption{$2$-opt neighborhood}
    \label{alg:2-opt}
    \begin{algorithmic} 
    \REQUIRE{$e_{i, j}$ is the edge connecting the vertice $i$ and $j$ in route $r_k$.}
    \ENSURE{Route $r_k$, cost $c_{i, j}$ between the vertice $i$ and $j$.}
    \WHILE{True}
    \STATE{count = 0}
    \STATE{$n = \text{len}(r_k)$}
    \FOR{$i = 1:n - 2$}
    \FOR{$j = i + 2:n$}
    \IF{$c_{i, i+1} + c_{j, j+1} > c_{i, j} + c_{i+1, j+1}$}
    \STATE{Cut $e_{i, i+1}$ and $e_{j, j+1}$}
    \STATE{Reconnect $e_{i, j}$ and $e_{i+1, j+1}$}
    \STATE{$\text{count} \, \mathrm{+}{=} \, 1$}
    \ENDIF
    \ENDFOR
    \ENDFOR
    \IF{$\text{count} = 0$}
    \STATE{break}
    \ENDIF
    \ENDWHILE
    \RETURN{$r_k$}
    \end{algorithmic}
\end{algorithm}
\label{fig:alg_2opt}
\end{figure}
\subsection{The Triangular Modification}
The route obtained from the Table~\ref{table:method_qubo_table} infringes on the rule $2)$. 
Accordingly, we implemented a triangular modification of the solution as soon as it was generated.
This postprocessing method is followed by a $2$-opt neighborhood search.
The example route in Figure~\ref{fig:post_processing_routes}(a) includes subtour $[v_3 \rightarrow v_0 \rightarrow v_4]$ where the vehicle returns to the depot and departs from it once again. 
Because the distance matrix satisfies the triangle inequality, we amended this subpath to $[v_3 \rightarrow v_4]$ which is certainly shorter than the previous one. 
The time complexity of the postprocessing method is estimated to be $\mathcal{O}(T)$.
Figure~\ref{fig:post_processing_routes}(b) shows the modified route. 
Note that a $2$-opt neighborhood search was also executed. 

Just as we apply the $2$-opt neighborhood search to CG, this triangular modification is called when the initial routes are generated and every time the pricing subproblems are solved. 
In contrast, prior to the $2$-opt neighborhood search, a triangular modification was employed in the event of the final solution acquisition regarding AE.
\begin{table}[t]
    \centering
    \caption{The problem setting.}
    \begin{tabular}{ccccccccccc}
         \hline
         $N$ & $T$ & $U$ & $d_\text{max}$ & $Q$ & $\sigma_\text{d}$ & $n_\text{sp}$ & $t_\text{sp}$ / s & $n_\text{AE}$ & $t_\text{AE}$ / s \\ \hline
         $40$ & $10$ & $6$ & $1$ & $26$ & $0.00$ & $5$ & $1$ & $30$ & $600$\\ 
         $40$ & $10$ & $6$ & $10$ & $60$ & $2.90$ & $6$ & $1$ & $36$ & $600$\\
         $40$ & $10$ & $6$ & $30$ & $122$ & $9.15$ & $7$ & $1$ & $42$ & $600$\\
         $40$ & $10$ & $6$ & $50$ & $169$ & $13.9$ & $8$ & $1$ & $48$ & $600$\\
         $40$ & $10$ & $6$ & $100$ & $365$ & $27.3$ & $9$ & $1$ & $54$ & $600$\\ \hline
    \end{tabular}
    \label{tab:problem_setting}
\end{table}
\subsection{Limited Column Generation}
Because the objective function of the pricing subproblem is given by the Equation~\eqref{eq:sp_obj}, the customer $v_i$, whose dual values are large, tends to be chosen to lower the objective value when using column generation. 
This potentially entraps the entire algorithm in a local optimum. 
Therefore, we propose fixing some variables in accordance with the following rule: customers in the $i$th iteration of the obtained route must not be visited in the $(i+1)$th iteration.
More specifically speaking, let $\tilde{V}$ be the vertex set of the customers selected at the $i$th iteration.
We introduce the variable fixation shown below:
\begin{equation}
\label{eq:fix_qubo_vars}
    q_{t, j} = 0 \qquad ^\forall j \in \tilde{V}, \quad 1 \leq t \leq T.
\end{equation}
This fixation allows us to explore the solution space globally because the generated columns should have little similarity to each other while simultaneously reducing the number of logical spins. 
The time complexity of this fixation can be estimated as $\mathcal{O}(|\tilde{V}|T)$.
Most importantly, the essence of this method is to remove the overlapping of the columns that are to be generated. 
Consequently, this fixation method can be applied to other optimization problems with inequality constraints if it can be formulated as a set cover problem.
We call this method Limited Column Generation (Limited CG).
Note that we performed the $2$-opt neighborhood search and triangular modification in the same manner as for CG. 

%% file: results.tex
\section{Numerical Results}
\label{Sec:results}
\noindent We compared the three different methodologies explained in the last section. 
To achieve this, we parameterized the difficulty of the inequality constraints by designating the maximum value of demand $d_\text{max}$. 
Demand $d_i$ for each customer is uniformly generated between closed intervals $[1, d_\text{max}]$.
Note that when $d_\text{max} = 1$, CVRP is equivalent to a pure VRP because each demand satisfies $d_i = 1$ for all the customers $v_i \in V$.
However, as the value of $d_\text{max}$ increases, the demand distribution becomes more complicated, leading to a challenging energy landscape.
Furthermore, the location of each customer is generated uniformly within a closed interval $[0, 5]$.

Table~\ref{tab:problem_setting} lists the parameters utilized to execute CG, Limited CG, and AE. 
$\sigma_\text{d}$ in Table~\ref{tab:problem_setting} refers to the standard deviation of the demand. 
The numbers of slack variables employed to express the inequality constraints are denoted by $n_\text{sp}$ and $n_\text{AE}$. 
The value of $n_\text{sp}$ corresponds to when CG and Limited CG are executed.
The value of $n_\text{AE}$ is the notation for AE. 
\begin{figure*}[t]
\centering
\subfloat[]{\includegraphics[keepaspectratio, scale=0.45]{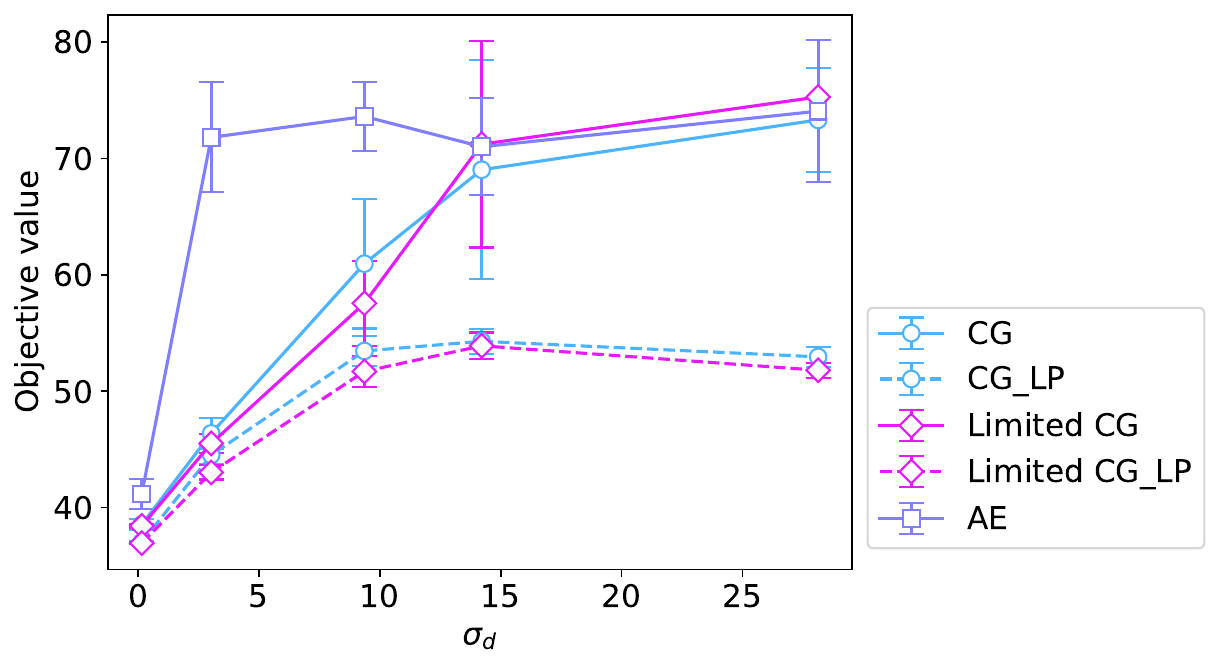}%
\label{fig:mainresult_a}}
\hfil
\subfloat[]{\includegraphics[keepaspectratio, scale=0.45]{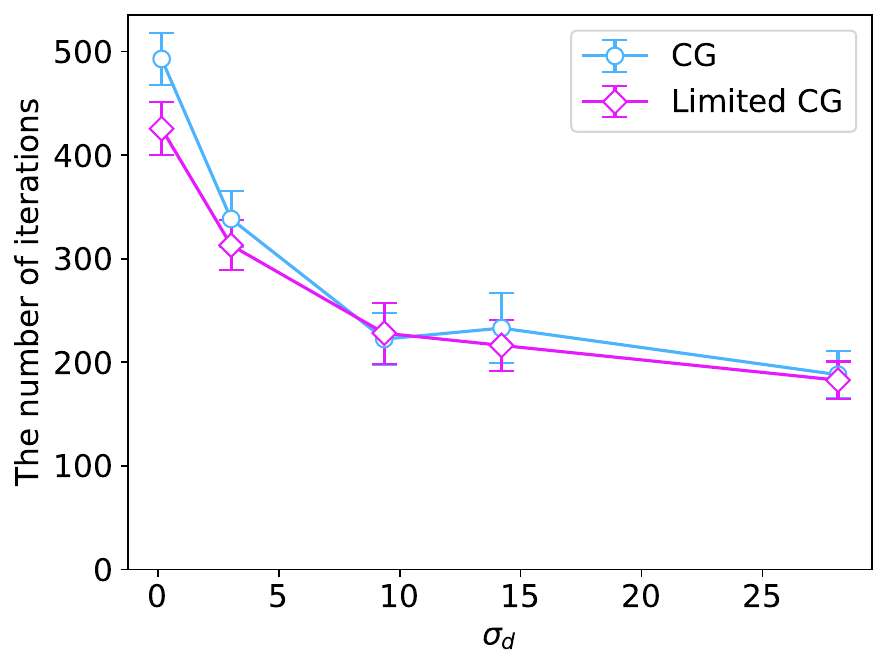}}%
\label{fig:mainresult_b}
\caption{Dependence of objective values on $\sigma_\text{d}$ when $N=40$, $U=6$, $T=10$. The blue circles, pink rhombuses, and purple squares denote CG, Limited CG, and AE, respectively. The error bars indicate the standard deviation of ten instances. (a) The solid line represents the final objective values. By contrast, the dotted line presents the relaxed objective values. (b) The number of iterations CG and Limited CG took to converge for each $\sigma_\text{d}$. }
\label{fig:result_difficulty_inequality}
\end{figure*}
\begin{figure*}[!t]
\centering
\subfloat[]{\includegraphics[width=2.15in]{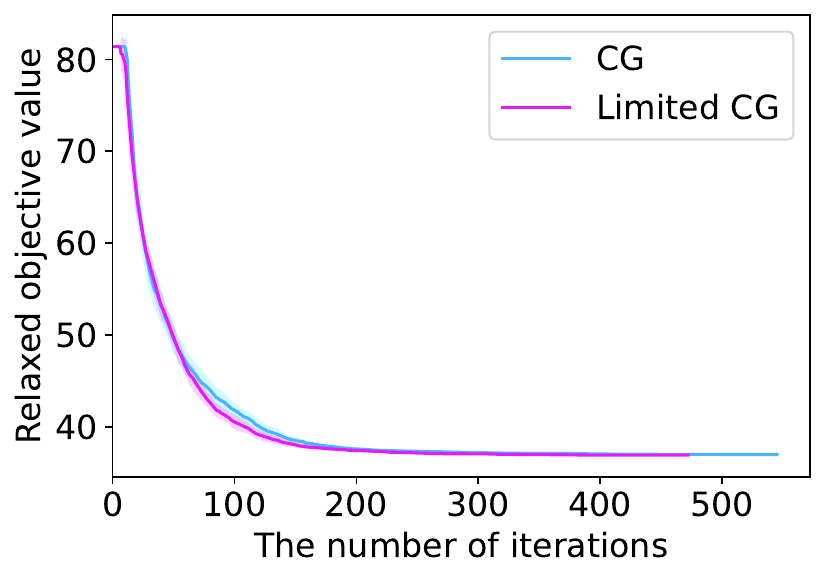}%
\label{fig:lp_objval_a}}
\hfil
\subfloat[]{\includegraphics[width=2.2in]{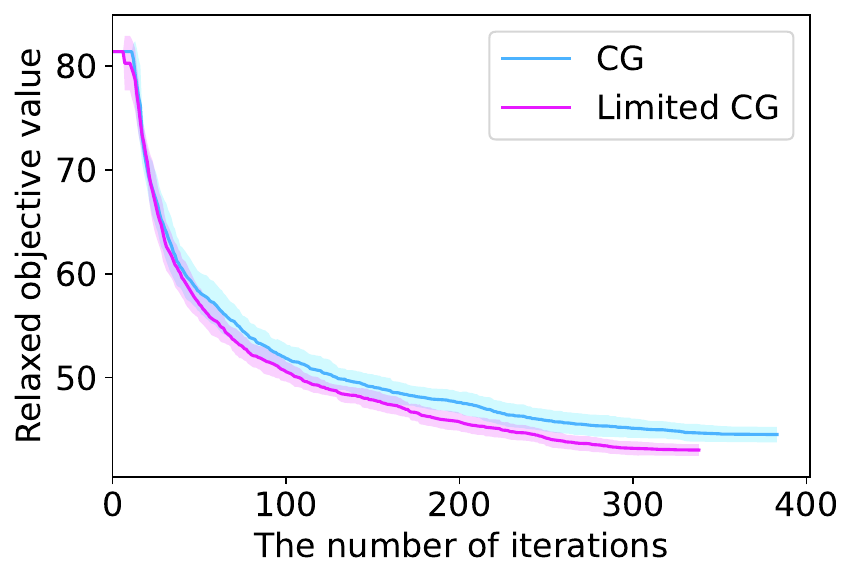}}%
\label{fig:lp_objval_b}
\hfil
\subfloat[]{\includegraphics[width=2.2in]{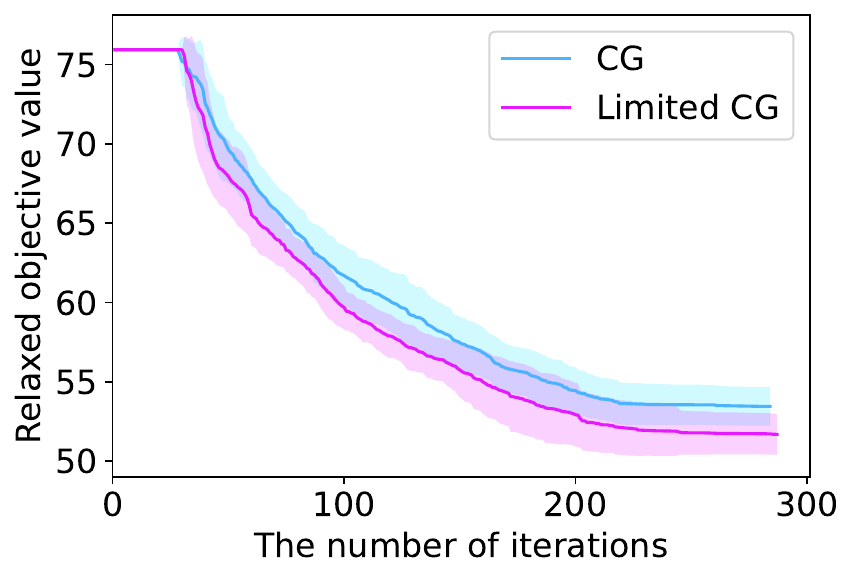}}%
\label{fig:lp_objval_c} \\
\subfloat[]{\includegraphics[width=2.2in]{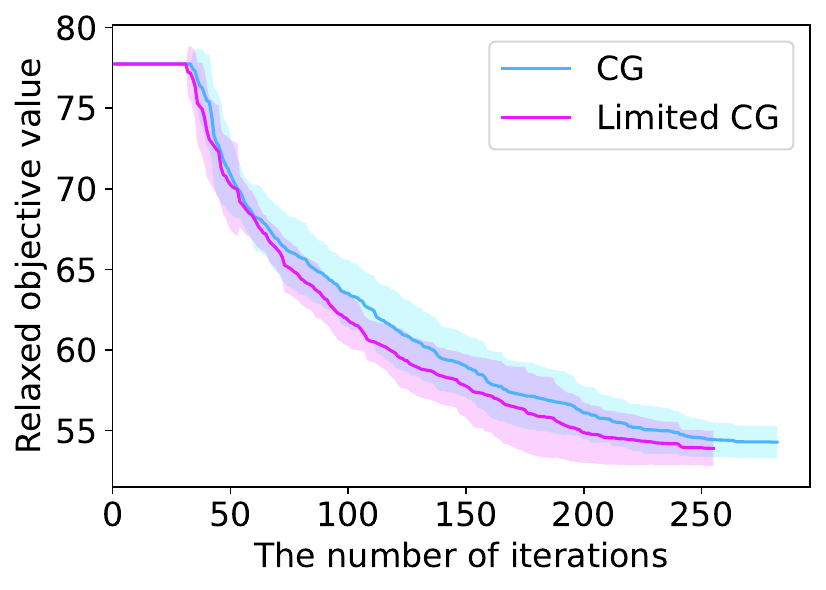}}%
\label{fig:lp_objval_d}
\hfil
\subfloat[]{\includegraphics[width=2.2in]{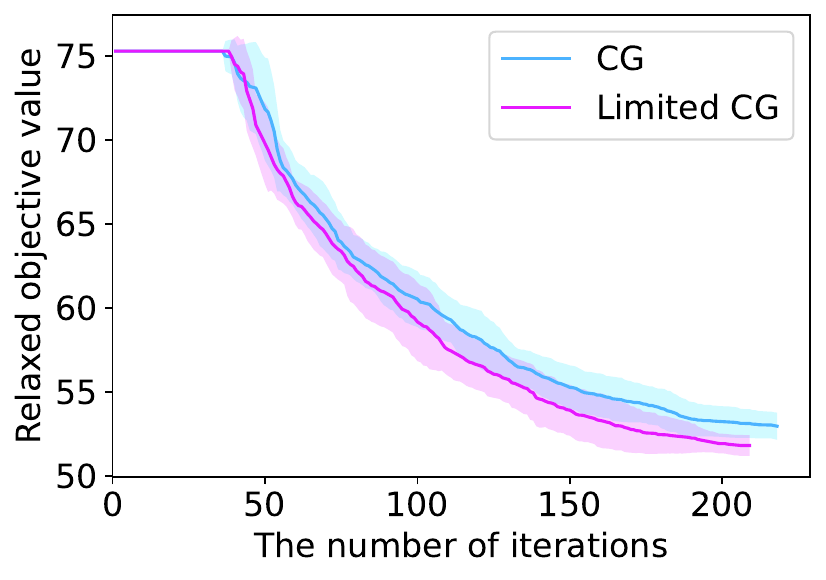}}%
\label{fig:lp_objval_e}
\caption{The time evolution of objective value of the restricted master problem. The color bar indicates the standard deviation. (a) $d_\text{max}=1$. (b) $d_\text{max}=10$. (c) $d_\text{max}=30$. (d) $d_\text{max}=50$. (e) $d_\text{max}=100$.}
\label{fig:lp_objval_iterations}
\end{figure*}
We discovered that this number of slack variables results from the use of binary encoding was used to transform an inequality constraint into an equality constraint.
Furthermore, note that $n_\text{sp}$ and $n_\text{AE}$ have the following relationship based on the difference in the formulation of the QUBO model: 
\begin{equation}
\label{fig:relationship_slack_vars}
    n_\text{AE} = U n_\text{sp}.
\end{equation}
In addition, $t_\text{sp}$ denotes the timeout for the use of Fixstars Amplify AE in executing CG and Limited CG. 
$t_\text{AE}$ is here again the notation for AE. 
To be sure, we solved the pricing subproblem once again with one second added to the current timeout if a route with a negative reduced cost could not be obtained.
The number of this additional explorations was limited to ten times.
To ensure fairness, the total computation time of CG and Limited CG was set to $600$ s, which was equal to the AE computation time.
\subsection{Analysis Based on Difficulty of Inequality Constraints}
Ten instances of CVRP were solved for each problem setting, as shown in Table~\ref{tab:problem_setting}.
Figure~\ref{fig:result_difficulty_inequality} shows the dependence of the objective values and the number of iterations on $\sigma_\text{d}$.
As shown in Figure~\ref{fig:result_difficulty_inequality}(a), even AE could indeed efficiently solve the problem when $d_\text{max}$ is equal to $1$, which is equivalent to a pure VRP in this case.
However, CG and Limited CG were observed to be superior to AE when $d_\text{max}$ increased to $10$ and $30$.
When $d_\text{max}$ is $50$ and $100$, instance problems become excessively challenging to the extent that CG and Limited CG are unable to obtain better solutions than AE.
Although the objective values for integer variables between CG and Limited CG were not significantly different from each other, we confirmed that the relaxed objective value acquired by Limited CG was superior to that of CG.
This is because Limited CG overcame the issue of CG in which a feasible set of routes was rarely obtained by eliminating the overlap of routes. 
Figure~\ref{fig:result_difficulty_inequality}(b) shows the number of iterations that both CG and Limited CG took. 
There was no discernible enhancement in the speed at which each method reached convergence, whereas Figure~\ref{fig:result_difficulty_inequality}(b) indicates that Limited CG can achieve better results without wasting further computational resources.

Figure~\ref{fig:lp_objval_iterations} illustrates the time evolution of the objective value for a restricted master problem. 
When $d_\text{max}=1$, the two lines almost overlap and we cannot observe the superiority of Limited CG to CG.
Nevertheless, as $d_\text{max}$ increases, Limited CG can converge to a better lower bound at a faster rate than CG. 
As previously stated, obtaining a good lower bound is of paramount importance because the objective value gap between the original integer problem and its linear relaxation still exists when $N$ is finite. 
\subsection{The range of effectiveness for CG and Limited CG.}
When $\sigma_\text{d}$ was within $0.156 \leq \sigma_\text{d} \leq 9.36$, we confirmed the superiority of annealing-assisted column generation methods (CG and Limited CG) compared to the naive method (AE) in the last subsection. 
Further investigations were conducted to ascertain whether these annealing-assisted column generation methods were effective in resolving these problems.
In this context, we introduce another evaluation factor $\bar{N}=N/U$, which is the average number of customers per vehicle. 
Table~\ref{tab:problem_setting_effective_domain} presents the problem settings for the additional simulations. 
Owing to the difference in the model formulations, the performance of AE should be strongly affected by $\bar{N}$.
As shown in Table~\ref{tab:problem_setting_effective_domain}, the number of slack variables $n_\text{AE}$ increases for a small $\bar{N}$ whereas $n_\text{sp}$ remains the same.
\begin{table}[t]
    \centering
    \caption{The problem setting for the effective domain investigation (Fig.~\ref{fig:result_effective_domain}).}
    \begin{tabular}{ccccccccccc}
         \hline
         $N$ & $T$ & $U$ & $d_\text{max}$ & $Q$ & $\sigma_\text{d}$ & $n_\text{sp}$ & $t_\text{sp}$ / s & $n_\text{AE}$ & $t_\text{AE}$ / s \\ \hline
         $40$ & $25$ & $2$ & $30$ & $122$ & $9.15$ & $7$ & $1$ & $14$ & $600$\\ 
         $40$ & $16$ & $3$ & $30$ & $122$ & $9.15$ & $7$ & $1$ & $21$ & $600$\\
         $40$ & $14$ & $4$ & $30$ & $122$ & $9.15$ & $7$ & $1$ & $28$ & $600$\\
         $40$ & $12$ & $5$ & $30$ & $122$ & $9.15$ & $7$ & $1$ & $35$ & $600$\\
         $40$ & $10$ & $6$ & $30$ & $122$ & $9.15$ & $7$ & $1$ & $42$ & $600$\\ 
         $40$ & $8$ & $7$ & $30$ & $122$ & $9.15$ & $7$ & $1$ & $49$ & $600$\\ 
         \hline
    \end{tabular}
    \label{tab:problem_setting_effective_domain}
\end{table}
\begin{figure}[t]
    \centering
    \includegraphics[keepaspectratio, scale=0.5]{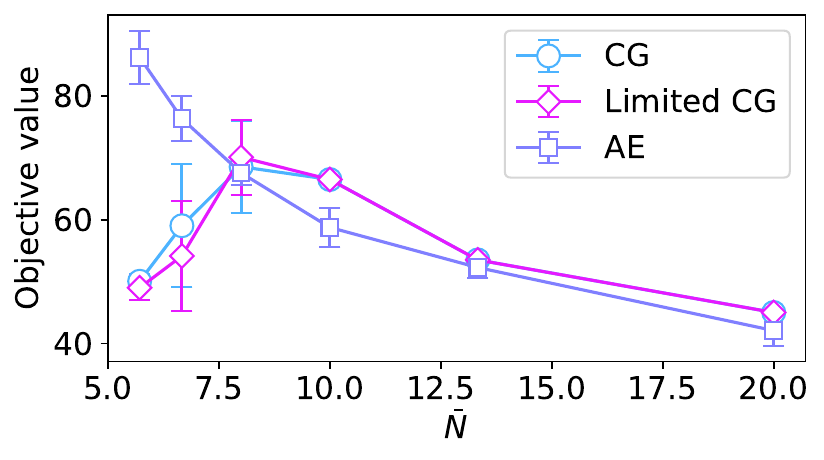}
    \caption{The range of effectiveness for CG and Limited CG. Error bars are used to indicate the standard deviation of ten instances.}
    \label{fig:result_effective_domain} 
\end{figure}
\begin{figure*}[t]
\centering
\subfloat[]{\includegraphics[keepaspectratio, scale=0.45]{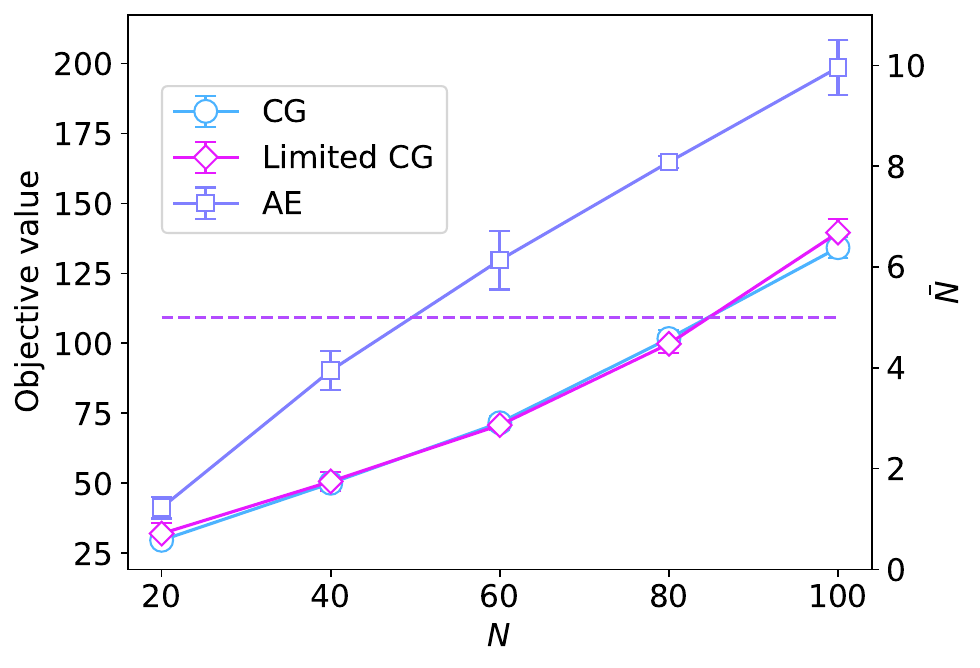}%
}
\hfil
\subfloat[]{\includegraphics[keepaspectratio, scale=0.45]{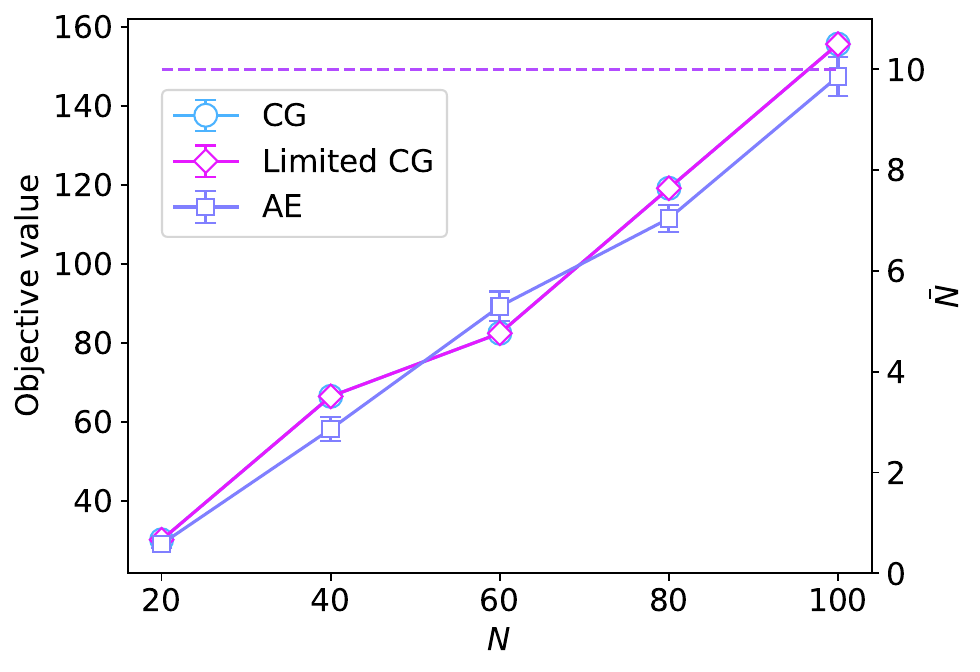}}%
% \label{fig:mainresult_b}
\caption{The scalability of each method for two different $\bar{N}$. We analyzed $5$ instances for every $N$. The left vertical axis presents the objective value, whereas the right vertical axis shows $\bar{N}$. In (a), $\bar{N}$ is set to $5$, while in (b), it is set to $\bar{N}=10$. }
\label{fig:result_scale}
\end{figure*}

Figure~\ref{fig:result_effective_domain} illustrates the dependence of the objective values on $\bar{N}$.
We revealed that CG and Limited CG are effective in regions in which $\bar{N}$ is small. 
This result does not contradict the finding that problems with many slack variables are difficult to resolve. 
Conversely, when $\bar{N}$ increases, the superiority of CG and Limited CG will disappears.

We also found that this relationship also applies to when the problem size scales.
Figure~\ref{fig:result_scale} shows the scalability of each method for two values of $\bar{N}$.
Figure~\ref{fig:result_scale}(a) shows the case where $\bar{N}=5$, while we can see the case where $\bar{N}$ is increased to $10$ from Figure~\ref{fig:result_scale}(b).
When $\bar{N}=5$ and $N=20$, the three methods were not significantly different because they had similar objective values. 
This result is corroborated by Equation~\eqref{eq:theorem_tightbound}, which indicates that column generation is an effective approach for large $N$.
As $N$ increases, the difference in the number of slack variables becomes significant. 
Figure~\ref{fig:result_scale}(b) shows the case in which $\bar{N}$ is $10$.
In this case, the column generation scheme had almost no supremacy over AE because $\bar{N}$ is relatively large, and the effects of the slack variables are diminished. 
This outcome is also attributable to the postprocessing algorithms' capacity to function affirmatively for AE, which has led to a significant improvement in the final path length. 
However, whether the ultimate result is favorable for CG and Limited CG does not highly depend on the postprocessing algorithms, but rather on the number of feasible combinations acquired until the final iteration. 
As seen earlier, the objective values of Limited CG are not significantly different from those of CG.
However, this result indicates that the column generation scheme is powerful enough to solve larger problems under specific conditions, offering a more effective alternative to the conventional naive approach. 

%% file: conclusion.tex
\section{Conclusion}
\label{Sec:conclusion}
\noindent In this study, we applied an annealing-assisted column generation method to the combinatorial optimization problems with inequality constraints. 
Such problems are typically challenging due to the hardware limitations of current Ising machines, with the abundance of decision and slack variables often leading to deteriorated final objective values.

Prior studies have addressed this challenge by introducing column generation and accelerating the solution process of the pricing subproblem using Ising machines~\cite{ossorio2022, da2023, hirama2023}. 
Further extending this approach to obtain a better lower bound, we introduced Limited CG, wherein we fixed the decision variables to eliminate the overlap between the generated columns.

Numerical simulations confirm that Limited CG can achieve a more robust lower bound, even as the complexity of each customer's demand increases.
Because finite-size problems are prevalent under practical scenarios, obtaining such a good lower bound is of paramount importance.
Additionally, we demonstrate that the average number of customers per vehicle $\bar{N}$ influences the efficacy of the column generation scheme, with annealing-assisted column generation methods proving effective for problems with small $\bar{N}$ , although this effectiveness diminishes as $\bar{N}$ increases.

Because our proposed method overlooks priority determined by dual solutions, future investigations should explore the limitations of overlap elimination.
In addition, the effectiveness of our method must be further assessed on benchmark problems.
Finally, given the considerable objective value gap between the integer problem and its linear relaxation in finite-size problems, we should develop a method for converting real solutions into integer solutions, ensuring this gap is effectively minimized.

%% file: acknowledgments.tex
\section*{Acknowledgments}
The authors wish to express their gratitude to the World Premier International Research Center Initiative (WPI), MEXT, Japan, for their support of the Human Biology-Microbiome-Quantum Research Center (Bio2Q).